\documentclass[superscriptaddress,showpacs,amssymb,10pt,reprint,aps,prd,longbibliography,nofootinbib,floatfix]{revtex4-1}

\usepackage{graphicx,epsfig,amssymb}
\usepackage{amsmath,amsfonts, times}
\usepackage{bm}

\usepackage{epstopdf}
\usepackage[linktocpage,colorlinks]{hyperref}
\usepackage[caption=false]{subfig}
\usepackage[usenames]{color}
\usepackage{natbib}
\usepackage{soul}
\usepackage[utf8x]{inputenc}
\usepackage[usenames]{color}
\usepackage{booktabs}
\usepackage{threeparttable}

\newcommand{\arccot}{\mathrm{arccot}\,}
\begin{document}
\newcommand{\beq}{\begin{equation}}
\newcommand{\eeq}{\end{equation}}
 \newcommand{\bqn}{\begin{eqnarray}}
 \newcommand{\eqn}{\end{eqnarray}}
 \newcommand{\nb}{\nonumber}
 \newcommand{\lb}{\label}
\newcommand{\PRL}{Phys. Rev. Lett.}
\newcommand{\PL}{Phys. Lett.}
\newcommand{\PR}{Phys. Rev.}
\newcommand{\PRD}{Phys. Rev. D.}
\newcommand{\CQG}{Class. Quantum Grav.}
\newcommand{\JCAP}{J. Cosmol. Astropart. Phys.}
\newcommand{\JHEP}{J. High. Energy. Phys.}
\title{Gravity-induced entanglement between two massive microscopic particles in curved spacetime: I. The Schwarzschild background}

	\author{Chi Zhang}
	\email{zhangchi3244@gmail.com}
	\affiliation{Department of Physics, Nanchang University, Nanchang, 330031, China}
	\affiliation{Center for Relativistic Astrophysics and High Energy Physics, Nanchang University, Nanchang,
330031, China}
		\author{Fu-Wen Shu}
	\email{shufuwen@ncu.edu.cn; Corresponding author}
	\affiliation{Department of Physics, Nanchang University, Nanchang, 330031, China}
	\affiliation{Center for Relativistic Astrophysics and High Energy Physics, Nanchang University, Nanchang,
330031, China}
\affiliation{GCAP-CASPER, Physics Department, Baylor University, Waco, Texas 76798-7316, USA}
\affiliation{Center for Gravitation and Cosmology, Yangzhou University, Yangzhou, China}

\begin{abstract}
The experiment involving the entanglement of two massive particles through gravitational fields has been devised to discern the quantum attributes of gravity. In this paper, we present a scheme to extend this experiment's applicability to more generalized curved spacetimes, with the objective of validating universal quantum gravity within broader contexts. Specifically, we direct our attention towards the quantum gravity induced entanglement of masses (QGEM) in astrophysical phenomena, such as particles traversing the interstellar medium. Notably, we ascertain that the gravitational field within curved spacetime can induce observable entanglement between particle pairs in both scenarios, even when dealing with particles significantly smaller than mesoscopic masses. Furthermore, we obtain the characteristic spectra of QGEM across diverse scenarios, shedding light on potential future experimental examinations. This approach not only establishes a more pronounced and extensive manifestation of the quantum influences of gravity compared to the original scheme but also opens avenues for prospective astronomical experiments. These experiments, aligned with our postulates, hold immense advantages and implications for the detection of quantum gravity and can be envisioned for future design.
	\end{abstract}

	\date{\today}
	
	\maketitle
		
\section{Introduction}
One of the significant cornerstones in the history of physics is the establishment of quantum field theory, which has successfully described the interaction of all relativistic fields except gravity. Despite decades of substantial efforts, the quantization of the gravitational field remains a challenging endeavor, leaving the comprehensive theory of quantum gravity still elusive. A significant barrier lies in the absence of experimental evidence supporting the quantum aspects of gravity, despite several proposed experimental designs aimed at detecting quantum gravity phenomenology \cite{r1}.

\begin{figure}\centering
\includegraphics[scale=0.3]{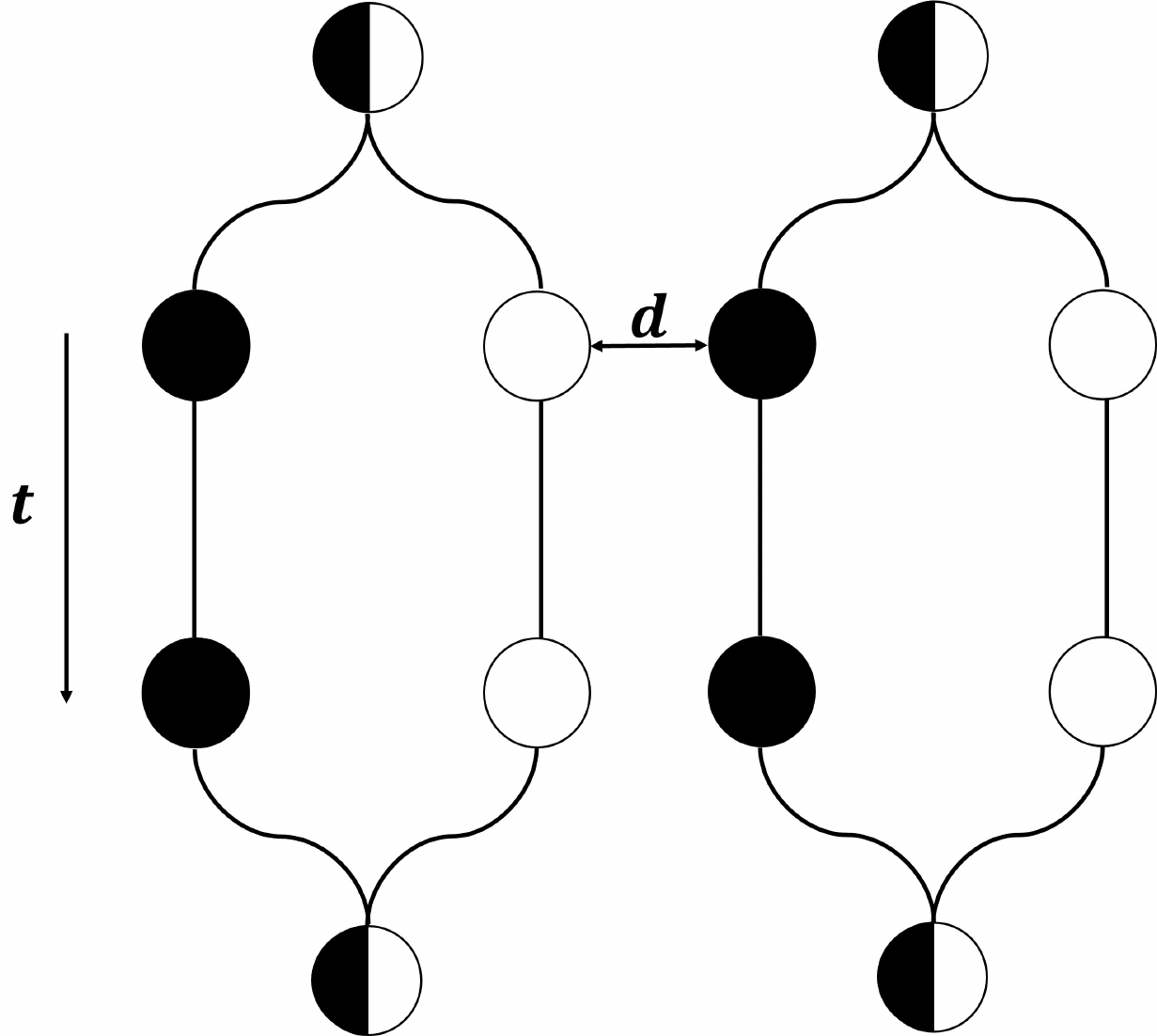}
\caption{The QGEM experiment setup: In the initial phase of this experiment, two mesoscopic particles, each initially in a spin superposition state, are placed a short distance apart. Subsequently, the application of an inhomogeneous magnetic field prompts each particle to assume a spatially split state contingent on its spin configuration. This step ensures a spin-dependent spatial separation. Following this, a coherent superposition of states is maintained for a certain duration, while keeping the separation distance fixed within each branch.
In the subsequent phase, the magnetic field is deactivated, and the spatially split states are realigned to regain their coherence. Finally, the experiment involves measuring spin correlations and calculating the entanglement witness. This analysis aims to determine if the system is indeed in an entangled state. Successful identification of entanglement would serve as confirmation of the quantum properties of gravity, consistent with the principles of entanglement theory.}\label{f11}
\end{figure}

Recently, the experiment concerning the quantum gravity induced entanglement of masses (QGEM), a laboratory-based proposal designed to measure quantum gravitational effects, was introduced by Bose et al. \cite{r2} and Marletto and Vedral \cite{r3}. This experiment involves the gravitational entanglement of two mesoscopic test particles. By observing the growth of entanglement between the particles in the QGEM scenario [see Fig. \ref{f11}], one can confirm the quantum nature of the gravitational field. Christodoulou and Rovelli \cite{r4} further extended this scheme to a generally covariant description, considering the effect as the quantum superposition of two distinct spacetime geometries along a particle's worldline. More recent advancements can be found in \cite{Capolupo:2019peg,Belenchia:2018szb,Carney:2018ofe,Marshman:2019sne,Buoninfante:2018xiw,Westphal:2020okx,Carlesso:2019cuh,Danielson:2021egj,Christodoulou:2022mkf,Cho:2021gvg,Matsumura:2020law,Miki:2020hvg,Howl:2023xtf,Yant:2023smr,Feng:2023krm,Schut:2023eux,Fragolino:2023agd,Li:2022yiy,Bose:2022uxe,He:2023hys}.

However, these previous studies assumed that massive particles were placed in an approximately flat spacetime (localized within the lab) and existed for short durations (seconds). In contrast, numerous astrophysical processes generate massive particles that interact over extended periods while traversing the universe, providing a natural setting for detecting the QGEM effect.

This paper proposes an innovative scheme to demonstrate gravity-induced entanglement in more general curved spacetimes, involving smaller-scale particles (microscopic when compared to the mesoscopic particles utilized in the QGEM setting). Specifically, we investigate the generation of entanglement between two particles in a Schwarzschild background to universally and convincingly test the quantum gravity experiment. Assuming that the particle pairs move along geodesic paths in each instance, the separation between each pair of trajectories will change due to geodesic deviation [see Fig. \ref{f10}]. Consequently, the proper time will vary between each pair of trajectories due to alterations in spacetime geometry. Intuitively, the proper time of the closest particles, which have the shortest spacelike distance between them, will experience the most significant increase. Building upon \cite{r4}, we compute the phase shift in each superposition state to determine the presence of entanglement. Remarkably, entanglement indeed emerges, illustrating the quantum gravity effect in general curved spacetime. We conduct a comprehensive analysis to explore the factors influencing the phase shift and their quantitative impact on entanglement.

However, a challenge remains in distinguishing whether the observed entanglement is generated by the quantum gravity effect of the particle pairs or by other processes during emission and propagation. To address this concern, we propose that QGEM during geodesic motion will exhibit a distinctive spectrum, as phase shifts occur across a series of geodesics. Analyzing the entangled patterns formed by various geodesics can assist in determining whether the entanglement originates from the gravitational field of nearby particles or from alternative sources.

The paper is structured as follows: In Sec. II, we provide a detailed description of our proposition for generating gravity-induced entanglement of microscopic massive particles in curved spacetime. In Sec. III, as an illustration, we consider a pair of particles moving within a globular cluster with a Schwarzschild-like metric and explore the influence of initial conditions on gravity-induced entanglement. We analyze two galactic models with different mass profiles, namely the dual pseudo-isothermal elliptical density (dPIE) profiles and the Navarro–Frenk–White (NFW) profile. In Sec. IV, we examine a more realistic scenario where the particles' geodesic trajectories deviate from the galaxy's center. Additionally, to uncover more features of QGEM, we investigate the entanglement witness as a function of the particles' kinetic energy. The paper concludes in the final section.

Throughout this paper, we adopt the natural units system, $c=G=1$, to simplify calculations. Physical quantities with units are provided in the SI system of units.

\section{Entanglement generation of microscopic particles in curved spacetime}
\subsection{The scheme}

In order to apply the QGEM scheme to astrophysics, its covariant description becomes crucial. The first covariant generalization of the QGEM scheme was undertaken by Christodoulou and Rovelli \cite{r4}. They also highlighted that gravity-induced entanglement arises from the superposition of spacetime geometry, leading to distinct changes in proper time across the four branches.

To delve into specifics, the particle pair is initially prepared in a superposition state of spin and spatial position:
\begin{equation}
\left|\Psi_i\right\rangle=\frac{1}{2}\Big( {\left| {LL } \right\rangle}+ {\left| {LR } \right\rangle}+{\left| {RL } \right\rangle}+{\left| {RR } \right\rangle}\Big)\otimes\left| {B} \right\rangle,
\end{equation}
where ${\left| {LL} \right\rangle}= \left| {\Psi_1^L} \right\rangle\otimes\left| {\Psi_2^L } \right\rangle$, and so forth. The notation $\left| {B } \right\rangle$ represents the quantum state of the background gravitational field.

Subsequently, in accordance with the hypothetical quantum superposition of distinct spacetimes \cite{r4}, the four branches will exhibit distinct time evolutions. The term ${\left| { RL } \right\rangle }$, characterized by the shortest separation, will accrue the maximum phase due to the rapid growth of inherent time, as expressed by
\begin{equation}
\phi = - \frac{{{m_0}\tau }}{\hbar } \approx - \frac{{{m_0}t}}{\hbar }\left( {1 - \frac{{{m_0}}}{R} - \frac{{{m_0}}}{d}} \right),
\end{equation}
where $m_0$ signifies the particle's mass.
Consequently, the central phase difference responsible for entanglement is represented by
\begin{equation}
\delta \phi = - \frac{{{m_0}\delta \tau }}{\hbar } = \frac{{{m_0}^2t}}{{\hbar d}}.
\end{equation}
Upon recombining the two components of the superposition, the final state, accounting for an overall phase factor, can be articulated as
\begin{equation}\label{fstate}
\left|\Psi_f\right\rangle=\frac{1}{2}\left( {\left| { LL} \right\rangle + \left| { LR } \right\rangle + {e^{i\delta \phi }}\left| { RL } \right\rangle + \left| { RR } \right\rangle } \right).
\end{equation}
Evidently, this state embodies entanglement of the spins of the two test masses, signifying that the gravitational field manifests as a quantum phenomenon.

However, this covariant description \cite{r4} remains confined to a flat space-time background. Furthermore, within the QGEM setting, two masses are situated at the mesoscopic scale and are subjected to a Stern-Gerlach setting, leading to a superposition of two components where each particle occupies different positions. However, this operation becomes challenging for astrophysical sources located far from Earth.

\begin{figure}
\centering
\includegraphics[scale=0.5]{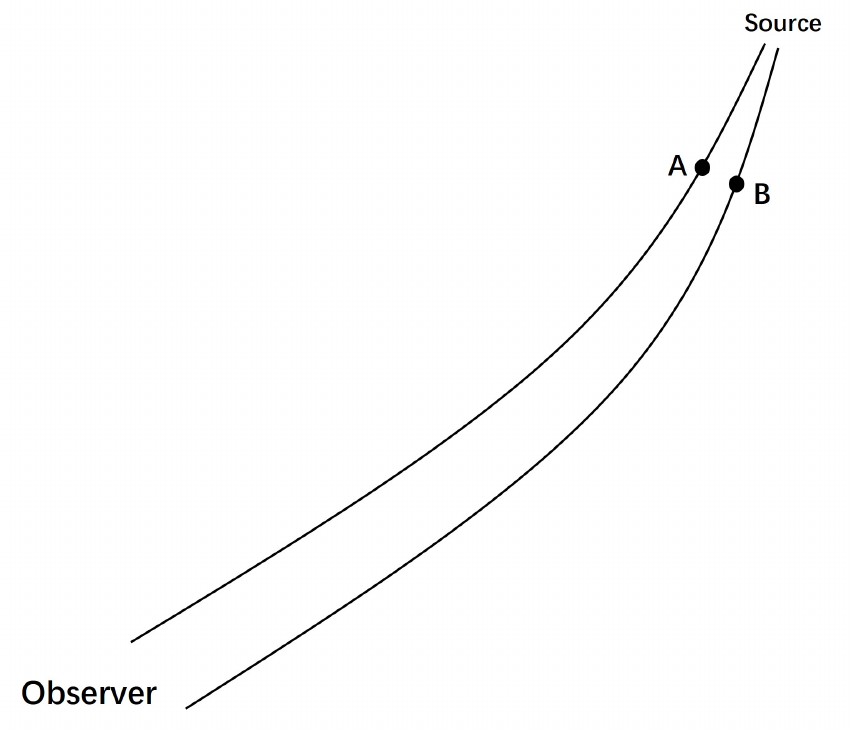}
\caption{Illustration of particle trajectories. The particles labeled as A and B are identical and microscopic in nature. They exist in a superposition state across two distinct spatial positions. The solid line delineates the geodesic trajectory of the particles within the curved spacetime.}\label{f10}
\end{figure}

To address these challenges, our approach deviates from the use of mesoscopic particles. In this study, we instead focus on microscopic massive particles, which had passed through a region with a magnetic field gradient during the propagation process in interstellar space and possessed a spatial superposition state. These microscopic particles freely fall within a generally curved spacetime, providing an opportunity to investigate gravity-induced entanglement. An overview of our general experimental setup is presented in Fig. \ref{f10}. 

Two identical microscopic particles, labeled A and B, both in the superposition state of two spatial positions as detailed in \cite{r2,r3}, traverse curved spacetime along their respective geodesic paths toward an Earth-based detector. Our main motivation is to employ a covariant methodology to comprehend the generation of entanglement throughout their journey.

In the reference frame of the moving particle, the particle's proper time is given by:
$\tau = \int {d\tau}$.
Let's now consider the scenario where another particle is in close proximity, moving alongside it but at a certain distance apart. The distance $d\left( \tau \right)$ varies over time due to geodesic deviation. In this context, we assume that the gravitational attraction between the particles is significantly weaker than the tidal force, allowing $d\left( \tau \right)$ to be primarily influenced by geodesic deviation. According to the equivalence principle in superposition spacetime \cite{Giacomini:2020ahk} and the Newtonian limit approximation, the proper time for the two particles in each branch can be expressed as:
\beq\label{32}
\tau = \int {\left( {1 - \frac{m}{{d\left( \tau \right)}} - \frac{m}{R}} \right)d\tau },
\eeq
where $R\sim \lambda_{c}$ ($\lambda_c$ representing the Compton wavelength of the particle) denotes the radius of each body. It should be much smaller than the distance,  $R\ll d(\tau)$, yet significantly larger than the Schwarzschild radius: $R  \gg r_s=2m$.
The phase difference between each branch is entirely attributed to the term:
\beq\label{33}
\delta \tau = - \int {\frac{m}{{d\left( \tau \right)}}d\tau },
\eeq
resulting in a phase change that can now be expressed as:
\beq\label{dphi}
\delta \phi = - \frac{{{m_0}\delta \tau }}{\hbar } = \frac{{{m_0}^2}}{\hbar }\int {\frac{1}{{d\left( \tau \right)}}d\tau },
\eeq
where $m_0$ denotes the static mass of the particles.

Subsequently, we will primarily focus on calculating the phase change in one typical astrophysical scenario: a spherically symmetric background, such as a globular cluster. It is important to note that, in this experimental design, for this order of analysis, the moving particles should be massive. This is because a relatively stationary observer cannot perceive the gravitational effects of a massless particle, such as a photon.

\subsection{Decoherence time}
Our above scheme a prior assumes that the particles are in their spatial superposition states.  However, in the complex background of the universe, various microscopic or mesoscopic particles, among which cosmic microwave background radiation (CMB) dominates, can potentially lead to decoherence. The decoherence rate of a neutral particle, initially in a superposition of two locations separated by $\Delta x$ and is immersed in a radiation heat bath at temperature $T$, can be described by the dipole approximation \cite{Joos:1984uk,Arrasmith:2017ogi}:
\beq\label{100}
\tau_\mathrm{D}^{-1}(\Delta x)=\left(16\frac{8!\zeta(9)}{9\pi}\right)\frac{\tilde{a}^6\Delta x^2(k_\mathrm{B}T)^9}{c^8\hbar^9},
\eeq
where ${\tilde a}$ represents the effective radius of the spherical particles. This formula holds when the radiation wavelength exceeds the particle radius and the superposition interval. For a superposition state with a particle mass of $10^{-25}$ kg and an upper limit of the superposition interval of $\Delta x$ in the CMB background, the lower limit of the decoherence time is estimated to be ${10^7} \times {\left( {\frac{m}{{\Delta x}}} \right)^2}s \sim {10^{19}} \times {\left( {\frac{m}{{\Delta x}}} \right)^2}s$. When this time scale greatly exceeds the Particle travel time, we are confidently assume that the superposition state is maintained throughout the entire particle movement. Of course, the scattering of impurity molecules in the universe and the thermal radiation of particles themselves will significantly increase the decoherence rate.

\section{Entanglement generation by globular galaxy}
Our example focuses on entanglement induced by the gravitational field within a cluster of galaxies, teeming with neutral massive particles and possessing a size substantial enough to generate a pronounced entanglement effect. To provide a comprehensive exploration, we investigate two distinct galactic models with differing mass profiles. The first model employs the dual pseudo-isothermal elliptical density (dPIE) profiles \cite{limousin,Eliasdottir:2007md}, which have been demonstrated to aptly describe the mass distributions of the brightest cluster galaxies (BCGs) \cite{r8}. These profiles find widespread use in lensing studies and deliver accurate fits to observed galaxies. The second model encompasses the Navarro–Frenk–White (NFW) profile \cite{nfw}, derived from N-body simulations, and widely employed for simulating dark matter (DM) halos within the $\Lambda$CDM universe.

\subsection{dPIE model}

In this subsection, let us envision an isolated galaxy cluster within the universe. For the sake of simplicity, we will exclusively consider the brightest cluster galaxy (BCG) component of the cluster. As previously indicated, the spherical dual pseudo-isothermal elliptical (dPIE) profiles prove especially fitting for characterizing the mass distributions of BCGs. These profiles are defined by their 3D-density \cite{limousin,Eliasdottir:2007md,r8}:
\begin{equation}\label{1}{\rho _{\text{dPIE}}}\left( r \right) = \frac{{{\rho _0}}}{{\left( {\frac{{{r^2}}}{{r_{\text{core}}^2}} + 1} \right)\left( {\frac{{{r^2}}}{{r_{\text{cut}}^2}} + 1} \right)}},\end{equation}
where $r$ is the distance from the center of the mass distribution, $r_{\text{core}}$ is the core radii and and $r_{\text{cut}}$ is the truncation radii with  $r_{\text{cut}}>r_{\text{core}}$. While $\rho_0$ is the central density, which is related to the 1D-central velocity dispersion, $\sigma_0$, by \cite{limousin}
\beq
\rho_0=\frac{\sigma_0}{2\pi G}\frac{r_{\text{cut}}+r_{\text{core}}}{r_{\text{core}}^2r_{\text{cut}}}.
\eeq

\begin{table}[h]\centering
\caption{Parameter choices of two different galaxy models}
\begin{tabular}{ccccc}\hline\hline
Galactic  & Model Parameters & Radius & ${\rho _0}$ & Mass\\
    Models               & (kpc)        &R (kpc)           &  (${M_ \odot } \cdot \text{kpc}^{ - 3}$)  & ($M_ \odot$) \\
                             \hline
dPIE                     & ${r_{\text{core}}} = 1.2,{r_{\text{cut}}} = 38$ &      2500             &   $ 2 \times {10^8}$ &  $2.07 \times {10^{11}}$    \\
NFW                     &     ${r_{s}} = 260$               &      2500             &   $ 644.5$           &  $2.07 \times {10^{11}}$   \\
 \hline\hline
\end{tabular}\label{table1}
\end{table}

Integrating Eq. \eqref{1}, one can obtain the total mass $m(r)$ enclosed by a sphere of radius $r$
\begin{equation}\label{2}\begin{split}\displaystyle m\left( r \right) &= 4\pi \mathop \smallint \nolimits_0^r {x^2}{\rho _{\text{dPIE}}}\left( x \right){\rm{d}}x\\
 \displaystyle &= \frac{{4\pi {\rho _0}r_{\text{core}}^2r_{\text{cut}}^2\left({r_{\text{cut}}}\arctan\left( {\frac{r}{{{r_{\text{cut}}}}}} \right)- {r_{\text{core}}}\arctan\left( {\frac{r}{{{r_{\text{core}}}}}} \right) \right)}}{{r_{\text{cut}}^2-r_{\text{core}}^2}},\end{split}\end{equation}
when $r<R$, where $R$ is the total radius of the BCG. Hence, the total mass of the galaxy is
\beq\label{M}
M = m(R),
\eeq
when $r>R$. The complete set of parameters utilized in the model adopted for this paper is enumerated in Table \ref{table1}. The parameters of the BCG model are grounded on the fitting conducted on the A383 cluster, as demonstrated in \cite{r8}. The parameter values for NFW, which will be employed in the subsequent subsection, also derive from the A383 dataset. However, to enable a meaningful comparison between the two models, suitable adjustments have been applied to ensure that both models possess identical radius and mass values.

With this mass profile in consideration, we are now prepared to present the metric for the background. Particularly, beyond the BCG, i.e., for $r>R$, the metric can be succinctly described by the Schwarzschild metric, with the mass specified in Equation \eqref{M} \cite{r9}.
\begin{equation}\label{3}
d{s^2} = \left( {1 - \frac{{2M}}{r}} \right)d{t^2} - \frac{dr^2}{{1 - \frac{{2M}}{r}}} - {r^2}\left( {d{\theta ^2} + {{\sin }^2}\theta {d^2}\varphi } \right).
\end{equation}
While for the interior of the BCG (i.e., $r<R$), the metric is given by [see Appendix \ref{inner metric}]
\begin{equation}\label{4}
d{s^2} = {e^{2A(r)}}d{t^2} - {e^{2B(r)}}d{r^2} - {r^2}\left( {d{\theta ^2} + {{\sin }^2}\theta {d^2}\varphi } \right),
\end{equation}
Here ${A(r)}$ is obtained by integrating $\mathop {A(r) = \smallint }\nolimits_0^r \frac{{m\left( x \right)}}{{{x^2}}}{\rm{d}}x$ and the explicit expression is given in Eq. \eqref{A5}, and $B(r)$ is given by
\beq\label{29}
B(r)= - \frac{1}{2}\ln \left(1 - \frac{{2m(r)}}{r}\right).
\eeq
For convenience in the following calculations, let us unify these two cases into one form like Eq. \eqref{4}, i.e.,
\begin{equation}\label{7}A(r) = -B(r)=\frac{1}{2}\ln \left( {1 - \frac{{2M}}{r}} \right)\ \ \text{for $r>R$}. \end{equation}

\begin{figure}\centering
\includegraphics[scale=0.4]{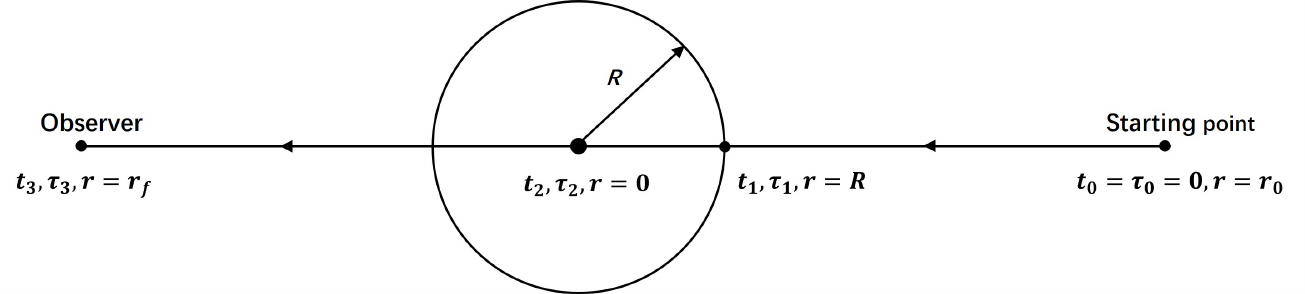}
\caption{Sketch of particle trajectory in galaxy cluster. It includes four different stages. The first stage refers to the trajectory from the starting point ($r=r_0, \varphi=\varphi_0$) to the boundary of the cluster ($r=R,\varphi=\varphi_0$). The second stage involves the boundary $r=R$ to the center of the cluster ($r=0$). The third stage is the reverse of the second one, namely, from ($r=0$) to ($r=R,\varphi=\varphi_0+\pi$). The last stage, which can be regarded as the reverse of the first stage, involves the geodesics from ($r=R,\varphi=\varphi_0+\pi$) to ($r=r_f,\varphi=\varphi_0+\pi$), the observer.}\label{f1}
\end{figure}

From now on, let us focus on one specific trajectory as shown in Fig. \ref{f1}. One of the particles starts at \({r_0}\), moving toward the center of the cluster and arrive the observer at \({r_f}\) finally. The whole trajectory follows a geodesic line through the center of the cluster.

It turns out the tetrad formalism is particularly convenient to discuss the geodesics. As to the present case, we construct the following four unity vectors,
\bqn\label{8}
{e_{(0)}}^{\mu} &= &e^{-A(r)}(\partial t)^{\mu},\  \
{e_{(1)}}^{\mu} = e^{-B(r)}(\partial r)^{\mu},\\
{e_{(2)}}^{\mu} &=& \frac{1}{r}(\partial \theta)^{\mu},\  \ \
{e_{(3)}}^{\mu} = \frac{1}{{r\sin \theta }}(\partial \varphi)^{\mu},\label{8b}
\eqn
so that they form an orthogonal base
\beq\label{normal}
g_{\mu\nu}{e_{(a)}}^{\mu}{e_{(b)}}^{\nu}=\eta_{(a)(b)},
\eeq
where $(a=0,1,2,3)$ and $\eta_{(a)(b)}=\text{diag}.(1,-1,-1,-1)$.

Recalling the geodesic equations in terms of four-velocity $u^{\mu}(\tau )$ is
\begin{equation}\label{fv}
{u^{\mu}} \circ ({e_{(a)} }_{\nu}{u^{\nu}}) + ({e^{(b)}}_{\nu}{u^{\nu}}){(d{e_{(b)} })_{\rho\sigma}}{u^{\rho}}{e_{(a)} }^{\sigma} = 0.
\end{equation}
For radial geodesics passing  through the center (the origin) as shown in Fig. \ref{f1}, the geodesic equations, after substituting Eqs. \eqref{4} and \eqref{8} into Eq. \eqref{fv}, reduce to
\begin{equation}\label{10}\begin{split}
&\displaystyle \ddot t + 2A'(r)\dot r\dot t = 0,\\
&\displaystyle {e^{2B(r)}}\ddot r + {e^{2B(r)}}B'(r){{\dot r}^2} + {e^{2A(r)}}A'(r){{\dot t}^2} = 0,
\end{split}\end{equation}
where the prime denotes differentiation with respect to $r$, while the dot represents  differentiation with respect to $\tau$. Note that the full trajectory can be split  into four stages as shown in Fig. \ref{f1}. The above equations \eqref{10} are only used for the first two stages. However,
using the spherically symmetric nature of the spacetime, the third and forth stages can be viewed as a reverse of the stages 2 and 1, respectively. Therefore, equations \eqref{10} can be used by redefining
\begin{equation}\label{11}\tilde t = 2{t_2} - t\left( {2{\tau _2} - \tau } \right),\tilde r = r\left( {2{\tau _2} - \tau } \right),\tilde \theta  = \theta ,\tilde \varphi  = \varphi  + \pi. \end{equation}

Our first goal is to solve these equations \eqref{10}, which are difficult to solve analytically.  We therefore find the numerical solutions. Without loss of generality, for later numerical computations, we assume that the particle has static mass ${m_0} = 10^{ - 25}$kg, which is a typical mass of a microscopic particle (with the Compton wavelength $\lambda_c\sim10^{-15}$ m).

\begin{figure}\centering
\subfloat[]{\includegraphics[scale=0.4]{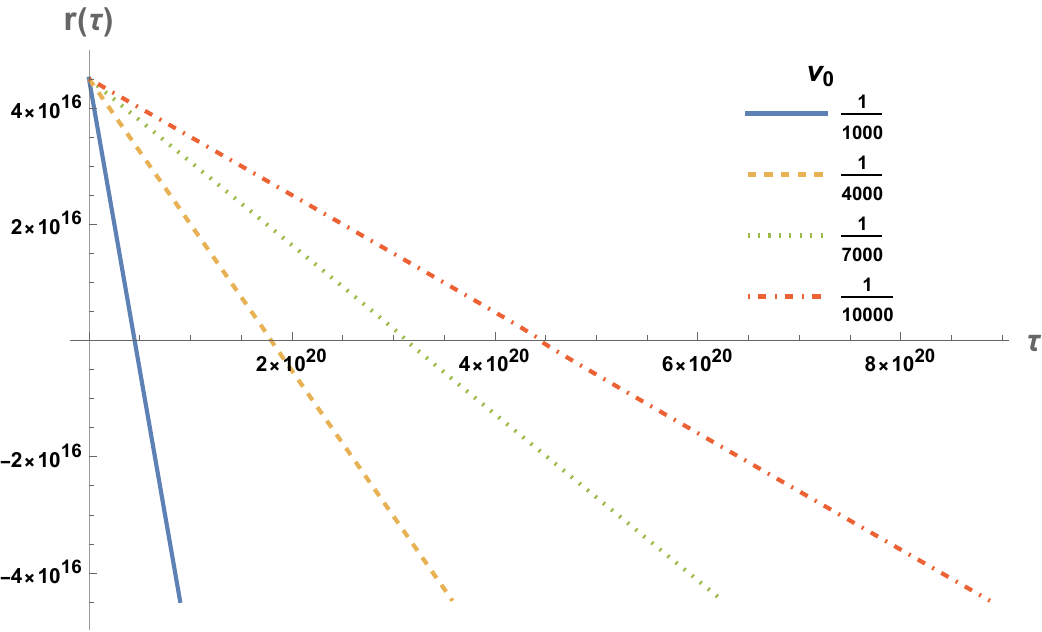}\label{f17a}}\\
\subfloat[]{\includegraphics[scale=0.4]{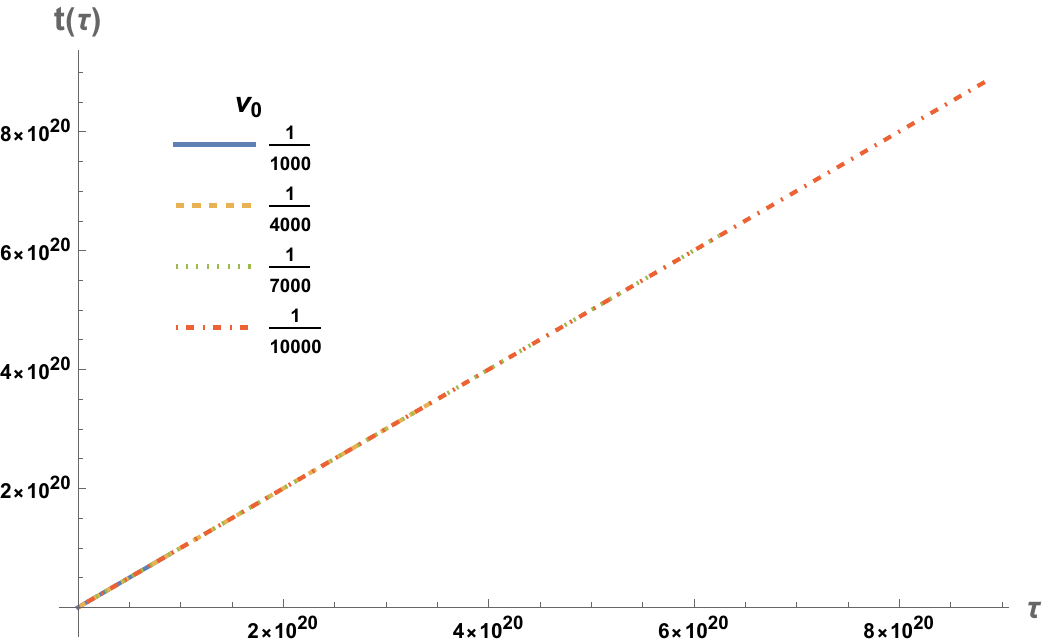}\label{f17b}}
\caption{Particle trajectories for different initial particle velocity $v_0$ with fixed $r_0= r_f\sim4.5\times 10^{16}$. (a) Relations between radial distance $r$ and the proper time $\tau$. Value of $r$ decreasing with $\tau$ indicates that the particle is propagating from the source to the cluster. Negative values of $r$ represent that the particle has passed through the center of the cluster. (b) Relations between temporal coordinate $t$ and the proper time $\tau$. They are insensitive to the initial velocity.}\label{f17}
\end{figure}

Fig. \ref{f17} shows a set of particle trajectories. The initial value for radial coordinate $r_0$  in our numerical calculations is taken to be $4.5\times 10^{16}$ (in natural unit\footnote{Translated to the International System of Units (SI system), $r_0=1.35\times 10^{22}\text{km}\sim 4.375\times 10^8 \text{pc}$.}), while the initial values for $\dot r(\tau=0)$ and $\dot t(\tau=0)$ can be transformed to the initial values of four velocity as
\beq\label{36}
u^{\mu}(\tau=0) = \gamma {e_{(0)}}^{\mu} - \gamma v_0 {e_{(1)}}^{\mu},
\eeq
where $\gamma  = \frac{1}{{\sqrt {1 - {v_0^2}} }}$ and $v_0$ is the initial velocity of the particle.

Fig. \eqref{f17a} illustrates that the value of $r$ decreases almost linearly with $\tau$, implying the particle's propagation from the source to the cluster. The steeper slope of the curve corresponds to a larger initial velocity. This correlation is reasonable since particles with higher $v_0$  will outpace those with lower $v_0$. In Fig. \eqref{f17b}, the curves of $t$ vs. $\tau$ largely overlap across the four different $v_0$ values, suggesting insensitivity to the initial velocity's magnitude.

Taking into account that the spacelike interval between two adjacent branches must exceed the Compton wavelength, we provide distinct initial conditions for the following scenarios. 

Now, let's proceed to calculate the geodesic deviation vectors, from which values of $d(\tau)$ can be deduced—this term plays a pivotal role in evaluating $\delta\phi$ as shown in Equation \eqref{dphi}. These vectors are determined through the solution of geodesic deviation equations in the tetrad formalism. Unlike the fixed tetrad used in Equation \eqref{8} for geodesics, the tetrad here must be parallelly transported along the geodesics \cite{r7}. In other words, ${{\tilde e}_{(a)}}{}^\mu {}_{;\nu }{v^\nu } = 0$, where $v^{\nu}$ is the tangent vector of the geodesics. Consequently, the axes' orientations remain fixed and non-rotating, as ascertained by local dynamical experiments. It turns out that the following selection of tetrads is appropriate
\bqn\label{12}
{{\tilde e}_{(0)}}{}^{\mu} &=& \dot t(\partial t)^{\mu} + \dot r(\partial r)^{\mu},\\
{{\tilde e}_{(1)}}{}^{\mu} &=& {e^{B - A}}\dot r(\partial t)^{\mu} + {e^{A - B}}\dot t(\partial r)^{\mu},\\
{{\tilde e}_{(2)}}{}^{\mu} &=& \frac{1}{r}(\partial \theta)^{\mu}, \ \  \
{{\tilde e}_{(3)}}{}^{\mu} = \frac{1}{{r\sin \left( \theta  \right)}}(\partial \varphi)^{\mu}.\label{e3}
\eqn

Note that in order for the above tetrad satisfying the orthogonal normalized condition \eqref{normal}, one extra condition should be imposed
\beq\label{nc}
e^{2A}\dot{t}^2-e^{2B}\dot{r}^2=1.
\eeq

The timelike basis ${\tilde e_{(0)}}{}^{\mu}$ is the geodesic four-velocity vector and the remaining three are spacelike basis pointing in different direction. All the four basis compose an orthogonally translational basis. In this tetrad basis the metric becomes just Minkowski metric.

Then according to the geodesic deviation equation in the above frames, we have
\beq\label{14}\frac{{{d^2}{w^{(a)} }}}{{d{\tau ^2}}} + {k^{(a)} }_{(b)} {w^{(b)} } = 0, \eeq
where $w^{(a)}$ is geodesic deviation vector in tetrad form
\beq
w^{(a)}={{\tilde e}^{(a)} }_{\mu}w^{\mu},
\eeq
and
\beq\label{35}{k^{(a)} }_{(b)}  =  - {R^{\mu}}_{\nu\rho\sigma}{{\tilde e}^{(a)} }_{\mu}{v^{\nu}}{v^{\rho}}{{\tilde e}_{(b)} }^{\sigma}.
\eeq

Substituting the metric \eqref{4}  and the tetrad \eqref{12} and \eqref{nc} into above definition, one finds that the nonvanishing components of ${k^{(a)} }_{(b)}$ are
\bqn\label{15}
{k^{(1)}}_{(1)} &=& {e^{ - 2B\left( r \right)}}\left( {A''\left( r \right) + A'{{\left( r \right)}^2} - A'\left( r \right)B'\left( r \right)} \right), \nonumber\\
\ \\
{k^{(2)}}_{(2)}& =& {k^{(3)}}_{(3)}= \frac{{B'\left( r \right){{\dot r}^2}}}{r} + \frac{{{e^{2A\left( r \right) - 2B\left( r \right)}}A'\left( r \right)}}{r}{{\dot t}^2}.\nonumber \label{k22}\\
\eqn

Without loss of generality, we assume that $w^{\mu}$ is orthogonal to ${{\tilde e}_{(0)} }{}^{\mu}$, such that we have $w^{(0)}=0$. Due to the spherical symmetry of the background,
the second and third components of $w^{(a)}$ are simply symmetric angular components.  Therefore, we can also choose suitable frame such that the initial value of $w^{(3)}$ is vanishing. Note that Eq. \eqref{14} is homogenous, hence if $\hat{w}^{(a)}$ is a solution of this equation, so is $\kappa\hat{w}^{(a)}$, where $\kappa$ is an arbitrary nonzero constant. Hence, once we initially have vanishing $w^{(3)}$ and $\dot{w}^{(3)}$, it will always be. In the end we are left with two nonzero components $w^{(1)}$ and $w^{(2)}$.

The total spacelike distance $d(\tau)$ between the pair of particles is
\beq\label{16}\begin{split}
d(\tau ) = \sqrt { {w^{(1)}}{{\left( \tau  \right)}^2} + {w^{(2)}}{{\left( \tau  \right)}^2}} \Delta l,\\
\tilde d(\tau ) = \sqrt { {{\tilde w}^{(1)}}{{\left( \tau  \right)}^2} + {{\tilde w}^{(2)}}{{\left( \tau  \right)}^2}} \Delta l,
\end{split}\eeq
where \textasciitilde{}  means the distance in the stages 3 and 4. $\Delta l$ is a constant and can be absorbed into $w^{(a)}$ in our numerical computations. Integrating over all four stages, one leads to the change of the proper time
 \begin{widetext}
\beq
\label{17}\delta \tau  =  - {m_0}\left( {\int_0^{{\tau _1}} {\frac{1}{{{d_1}\left( \tau  \right)}}} {\rm{d}}\tau  + \int_{{\tau _1}}^{{\tau _2}} {\frac{1}{{{d_2}\left( \tau  \right)}}} {\rm{d}}\tau  + \int_{{\tau _2}}^{{\tau _3}} {\frac{1}{{{{\tilde d}_3}\left( \tau  \right)}}} {\rm{d}}\tau  + \int_{{\tau _3}}^{{\tau _4}} {\frac{1}{{{{\tilde d}_4}\left( \tau  \right)}}} {\rm{d}}\tau } \right).
\eeq
 \end{widetext}
Once $\delta\tau$ is obtained, the change in phase can be known through
\beq\label{18}\delta \phi  =  - \frac{{{m_0}\delta \tau }}{\hbar }.\eeq
As a consequence, the key step is to get the values of $w^{(i)}$ and ${{\tilde w}^{(i)}}$ ($i=1,2$), which can be obtained by solving the geodesic deviation equations \eqref{14}, \eqref{15} and \eqref{k22}.

\begin{figure}\centering
\subfloat[]{\includegraphics[scale=0.45]{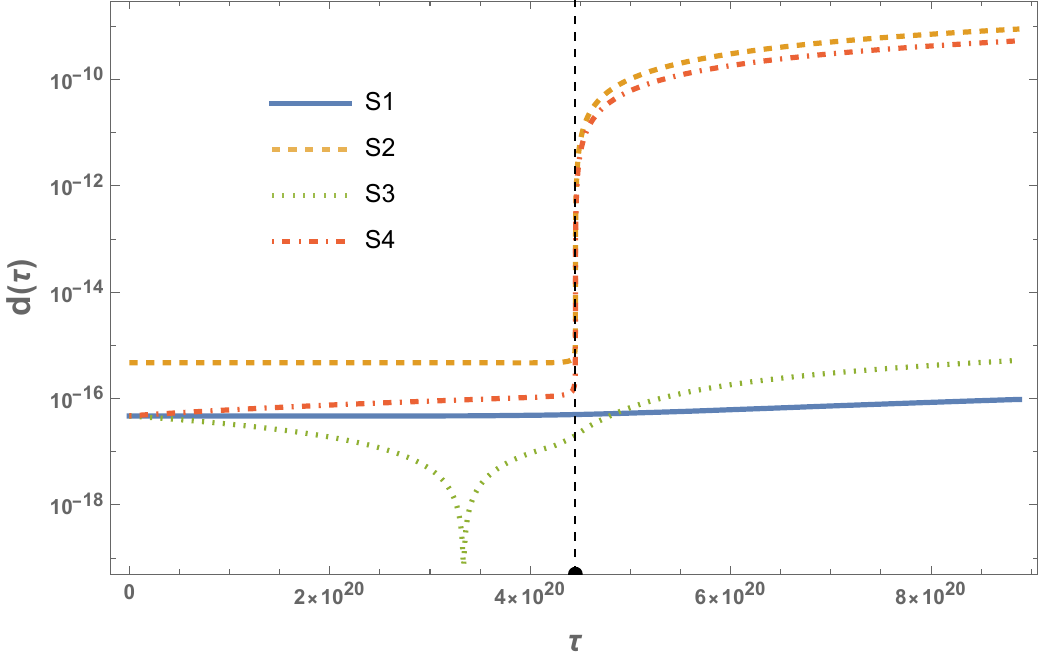}\label{5a}}\\
\subfloat[]{\includegraphics[scale=0.45]{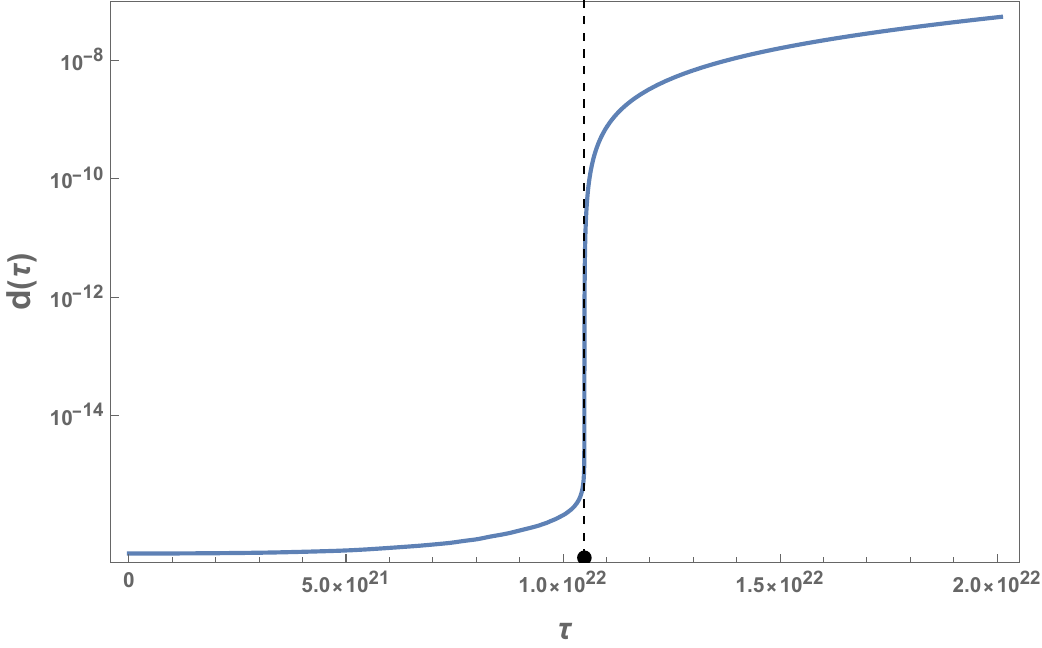}\label{5b}}
\caption{Spacelike separation $d(\tau)$ between particle pairs as a function of initial conditions. The black dashed line denotes the moment when the particle reaches the center of the cluster. (a) Overview of particles start for different initial conditions as listed in Table \ref{initial}. (b) Full view of initial value set S5. Particle's proper time integrated over the trajectory of this case is much longer than those of others.}\label{f18}
\end{figure}

In what follows let us turn to numerically solve the geodesic deviation equations \eqref{14}, \eqref{15} and \eqref{k22} under the metric ansatz Eqs. \eqref{4}-\eqref{7} and \eqref{A5},  and obtain $\delta \phi$ for appropriate initial conditions. Totally the initial conditions include: the initial geodesic deviations $w_0^{(1)}$ and $w_0^{(2)}$, the initial geodesic deviation velocities $\dot{w}_0^{(1)}$ and $\dot{w}_0^{(2)}$ measured in tetrad basis \eqref{12}, the initial radial coordinates ${r_0}$, and lastly the initial values for $\dot r(\tau=0)$ and $\dot t(\tau=0)$, which, again, can be transformed to the initial values of four velocity via \eqref{36} by setting suitable initial velocity $v_0$.  Totally there are six initial conditions to be assigned. However, for simplicity, in our following numerical simulations we will assume that $w_0^{(1)}=w_0^{(2)}$ and $\dot{w}_0^{(1)}=\dot{w}_0^{(2)}$. The full set of initial values which are adopted in the numerical simulations can be found in Table \ref{initial}. Note that throughout the paper, $z$ represents the redshift of the source\footnote{ The redshift $z$ can be calculated by using $$r(z)=r_0+r_f=\int^{0}_{z} \frac{-\mathrm{d}z}{H_0\sqrt{\Omega_{\Lambda}+\Omega_M(1+z)^3+\Omega_R(1+z)^4}},$$ where $\Omega_{\Lambda}$, $\Omega_M$ and $\Omega_R$ are the density parameters for dark energy, matter and radiation, respectively. $H_0$ is the Hubble constant. According to Plank\textbf{ 2018 }data \cite{Aghanim:2018eyx}: $\Omega_{\Lambda}=0.685$, $\Omega_m=0.315$, $\Omega_R=0$ and $H_0=67.4~km/s/\mathrm{M}pc$. Without loss of the generality, in the paper we take $r_f=4.5\times 10^{16}$ for simplicity.}.  It is not difficult to generalize the simulations to the cases where  $w_0^{(1)}\neq w_0^{(2)}$ and $\dot{w}_0^{(1)}\neq \dot{w}_0^{(2)}$. 

\begin{table}[h]\centering
\begin{threeparttable}
\caption{Sets of initial values for numerical simulations}
\begin{tabular}{ccccc}\hline\hline
Sets & redshift $z$& velocity $v_0$ & $w_0^{(1)}=w_0^{(2)}$ & $\dot{w}_0^{(1)}=\dot{w}_0^{(2)}$ \\
                             \hline
S1             & $0.2$  &      $10^{-4}$     & $ \frac13\times 10^{-16}$   & 0 \\
S2              & $0.2$  &     $10^{-4}$     &     $\frac13\times 10^{-15}$      & 0  \\
S3             & $0.2$  &     $10^{-4}$     &    $ \frac13\times 10^{-16} $     & $-10^{-37}$   \\
S4             & $0.2$  &     $10^{-4}$     &    $ \frac13\times 10^{-16}$       & $10^{-37}  $ \\
S5              & $0.2$  &     $0 $     &  $\frac13\times 10^{-16}$   & 0 \\
 \hline\hline
\end{tabular}\label{initial}
\begin{tablenotes}
\item[1] Note that this is in natural unit system. In SI system, $v_0$, $w_0^{(i)}$ and $\dot{w}_0^{(i)}$ should be multiplied by $c=3\times 10^8 \text{m/s}$. For example, for S1, we have ${v_0} = 3 \times {10^4}{\rm{m/s}} $, $w_0^{(1)}=w_0^{(2)}=10^{-8} \text{m}\ (\gg \lambda_c\sim 10^{-15}\text{m})$, and  $\dot{w}_0^{(1)}=\dot{w}_0^{(2)}=0$.
\end{tablenotes}
\end{threeparttable}
\end{table}

Fig. \ref{f18} displays geodesic deviation for various initial values, revealing significant changes in all space-like separations $d(\tau)$ near the center of the galaxy. Notably, the magnitude of $d$ directly affects the extent of change when approaching the galaxy center. In some cases, like curve S1 in Fig. \ref{5a}, where $d(\tau)$ is very small near the center, the turning point might not be clearly represented on the graph (indicated by the black dashed line in Fig. \ref{f18} as the particle reaches the center of the cluster). For particle pairs with a negative initial geodesic deviation velocity (particles initially moving towards each other), they initially converge and subsequently diverge. This behavior is consistently captured in our numerical results, as evidenced by the curve of set S3 in Fig. \ref{f18}, which first declines and nearly touches the zero point before passing through the center of the cluster, and then rises afterward.

\begin{figure*}\centering
\includegraphics[scale=0.11]{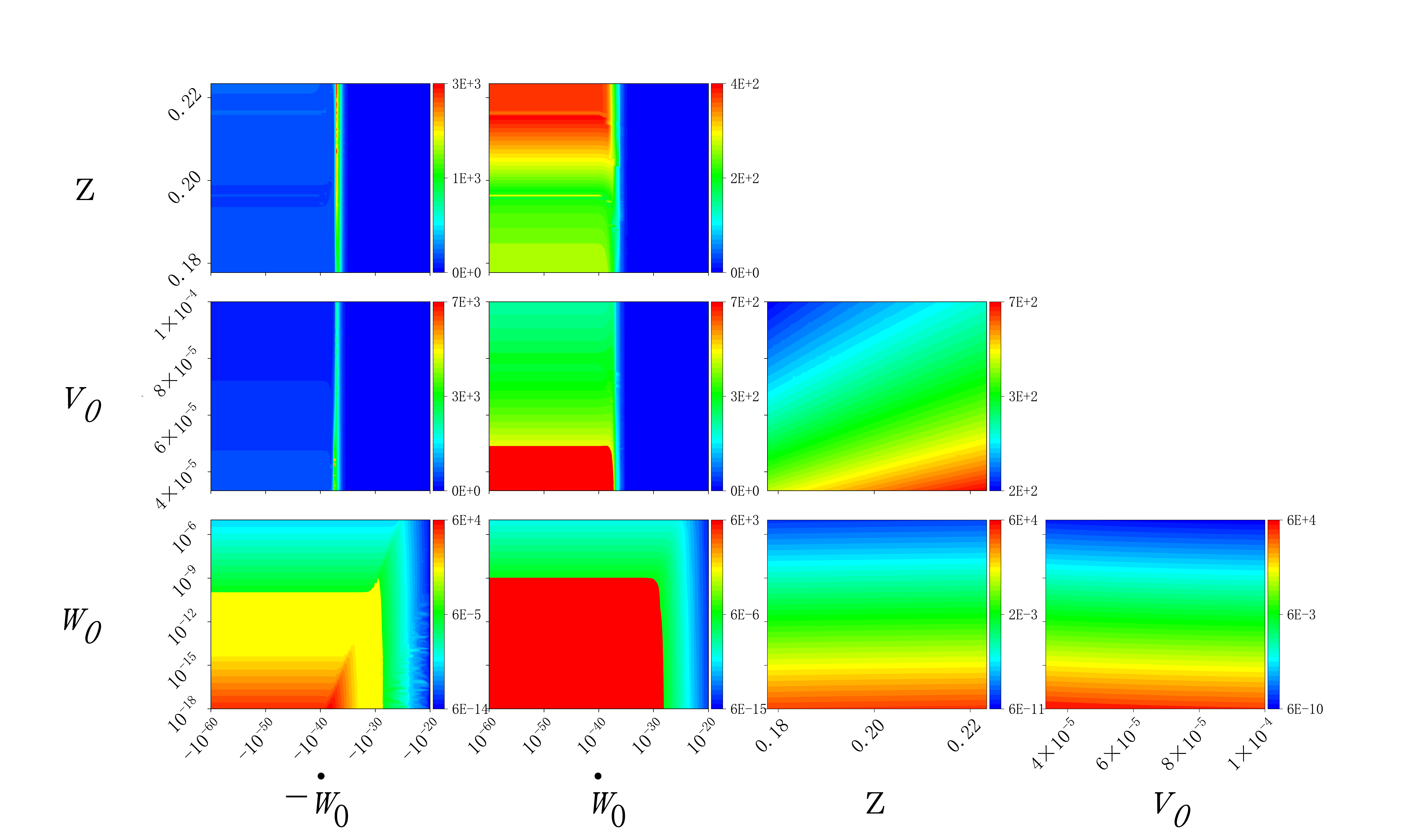}
\caption{Contour plot of phase changes $\delta \phi$ for initial value set S1 in Table \ref{initial}.  In these plots, we let two of four initial conditions vary in each subgraph, while the other two are fixed.  Different color denotes different values of the phase change. The following notations ${w_0^{(1)}} = {w_0^{(2)}}=w_0, {\dot w_0^{(1)}} = {\dot w_0^{(2)}}=\dot w_0$ have also been used.}\label{f12}
\end{figure*}

Phase changes $\delta\phi$ in different initial conditions are shown in Fig. \ref{f12}. In each subgraph, the shade of color represents the magnitude of the phase change varies with two of four initial values, with the remaining two values keep fixed.

From the left six pictures, it is obvious that the phase change is sensitive to the initial geodesic deviation velocity $\dot{w}_0$ no matter whether it is positive or negative.
More specially, from figures in the leftmost panel one can see that the phase change increases at the beginning with negative deviation velocity $-\dot{w}_0$ and then decreases. This is contrary to that of $d(\tau)$ as shown in Fig. \ref{5a}. This is reasonable because the separation distance $d(\tau)$ appearing in the denominator of Eq. \eqref{17}.  While figures in the second column show that $\delta\phi$ decreases with initial deviation velocity $\dot{w}_0$ monotonously, which is also consistent with the results as shown in Fig. \ref{f18}. In both cases, $\delta\phi$ increases monotonously with the redshift $z$, although it is relatively mild for the negative initial deviation velocity case. In contrast, the phase change decreases monotonously both with the initial particle velocity $v_0$ and initial deviation distance $w_0$ no matter whether the initial geodesic deviation velocity  is positive or negative.

On the other hand, three figures in the bottom right corner corresponds to the cases where the initial deviation velocity is 0, leaving dependence of $\delta\phi$ on $z, v_0$ and $w_0$, respectively. Specifically, the third column shows that for fixed $z$, $\delta\phi$ decreases with both $v_0$ and $w_0$ monotonously. In contrast, it increases with $z$ especially obvious for fixed $v_0$. This is because as $z$ increases, the proper time of the whole geodesic motion also increases, but the geodesic deviation is basically unaffected. The rightmost figure shows that for fixed $z$ and $v_0$, the phase change is inversely proportional to the initial geodesic deviation $w_0$. This is because the geodesic deviation equations are a set of homogeneous second-order differential equations and the spacelike interval at every moment is proportional to its initial value $w_0$.

\subsection{NFW model}

In this subsection let us consider the NFW model, which can be useful to observe the effects of the DM halos in the $\Lambda$CDM universe. The density profile and mass of the NFW model is \cite{r8}
 \beq\label{37}\begin{split}
 &{\rho _{\text{NFW}}}\left( r \right) = \frac{{{\rho _0}}}{{\frac{r}{{{r_s}}}{{\left( {1 + \frac{r}{{{r_s}}}} \right)}^2}}},\\
 &m\left( r \right) = 4\pi {\rho _0}{r_s}^3\left( { - \frac{r}{{r + {r_s}}} + \ln \left( {1 + \frac{r}{{{r_s}}}} \right)} \right),
 \end{split}\eeq
 where $\rho_0$ is the scaling density and $r_s$ is the scaling radius. These two parameters have relationship with the Virial mass of the halo through $M_{200}\propto \rho_0r_s^3$ and are related to the concentration parameter through $c = r_{200}/r_s$. In order to compare with the dPIE model, in this paper we set $\rho_0$ such that the total mass of the cluster is the same as the one of the dPIE model. The full list of the parameters can be found in Table \ref{table1}. Following the same procedures, one finds the metric parameters $A(r)$, $B(r)$ in \eqref{4} is
 \begin{widetext}
 \beq\label{38}\begin{split}
 &A(r)= 4\pi {\rho _0}{r_s}^3\left( {\frac{1}{{{r_s}}} + \frac{{\ln \left( {\frac{{{r_s}}}{{r + {r_s}}}} \right)}}{r}} \right) + c,\\
 &B(r)= - \frac{1}{2}\ln \left(1 - \frac{{2m(r)}}{r}\right),\\
 &c = \frac{1}{2}\ln \left( {1 - \frac{{8\pi {\rho _0}{r_s}^3\left( { - \frac{R}{{R + {r_s}}} + \ln \left( {\frac{{R + {r_s}}}{{{r_s}}}} \right)} \right)}}{R}} \right) - \frac{{4\pi {\rho _0}{r_s}^2\left( {{r_s}\ln \left( {\frac{{{r_s}}}{{R + {r_s}}}} \right) + R} \right)}}{R}.
 \end{split}\eeq
 \end{widetext}

\begin{figure}\centering
\includegraphics[scale=0.4]{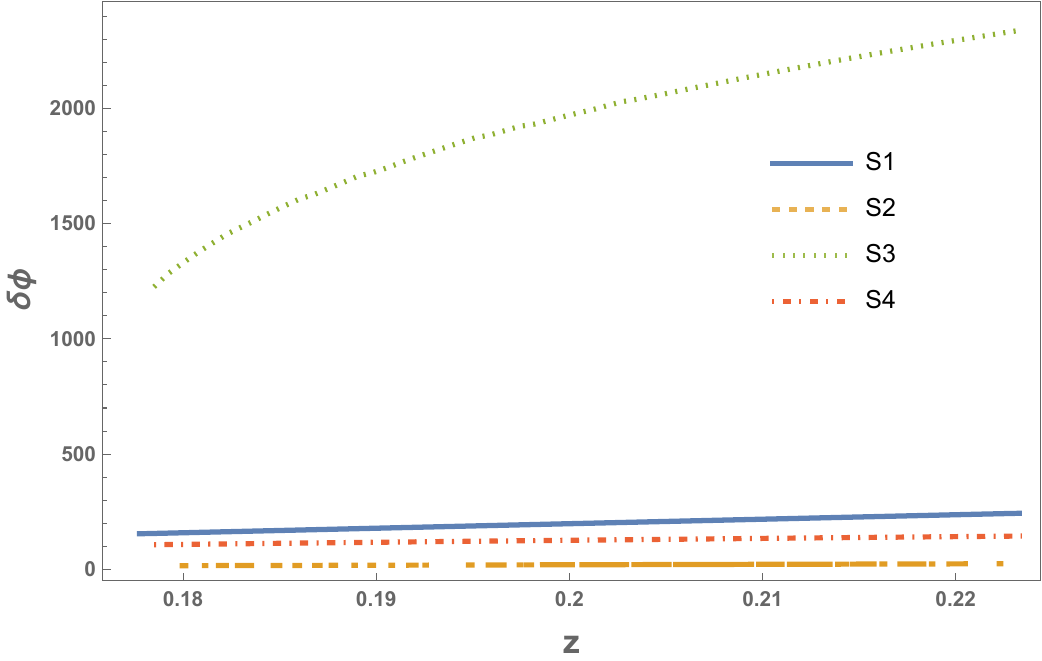}
\caption{The phase change $\delta \phi$ as a function of initial radial position (redshift) of the particles. The initial values of the other parameters are given in Table \ref{initial}.}\label{f14}
\end{figure}
\begin{figure}
\includegraphics[scale=0.4]{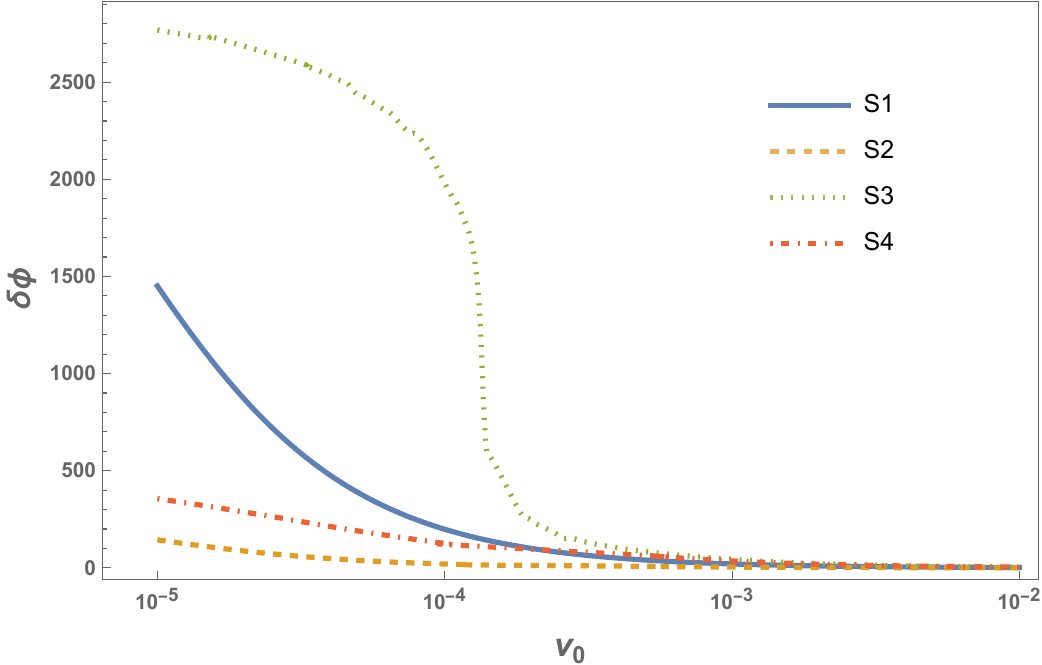}
\caption{The phase change $\delta \phi$ as a function of initial radial velocity $v_0$ of the particles. The initial values of the other parameters are given in Table \ref{initial}.}\label{f15}
\end{figure}
Following what we did in the last subsection, one can obtain phase change $\delta\phi$ for different initial conditions. 

Fig. \ref{f14} depicts the phase change $\delta\phi$ as a function of initial redshift, considering other initial values given in Table \ref{initial}. Notably, the plot exhibits a nearly linear growth of the phase change with increasing initial $z$. This behavior is attributed to the increase in the proper time of motion before passing through the center of the galaxy cluster, which serves as the integral variable, while the geodesic deviation after passing through the center remains almost unchanged. Consequently, the curve closely resembles a straight line.

In Fig.\ref{f15}, we plot the phase change $\delta \phi$ as a function of initial radial velocity $v_0$. One can see from this plot that the phase change decreases monotonously as increasing $v_0$. In addition, for small $v_0$, it decreases fast as $v_0$ increases and gradually the decrease becomes mildly. Finally it approaches zero gradually for large $v_0$. This is can be explained in the following way: the particle's total proper time integrated over the trajectory decreases precipitously as initial radial velocity $v_0$ goes from zero to non-zero. This has been already verified in the dPIE model  as shown in Fig. \ref{5b}.
\begin{figure}
\includegraphics[scale=0.6]{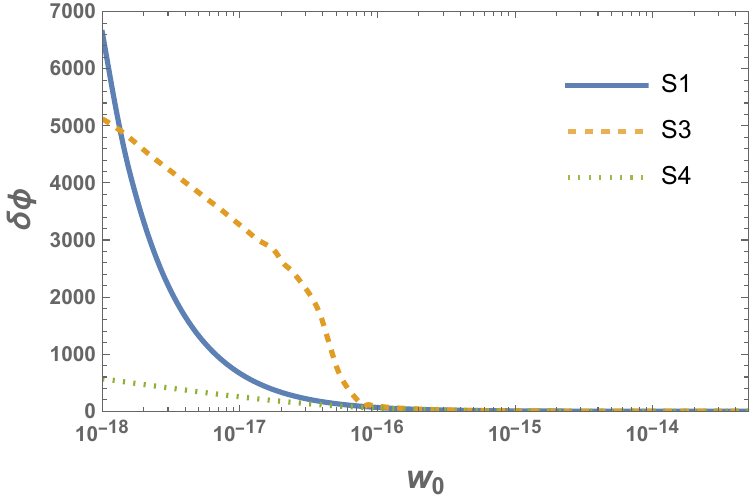}
\caption{The phase change $\delta \phi$ as a function of initial geodesic deviation $w_0$ of the particles. The initial values of the other parameters are given in Table \ref{initial}.}\label{f19}
\end{figure}

\begin{figure*}\centering
\subfloat[The initial geodesic deviation velocity is positive. ]{\includegraphics[scale=0.38]{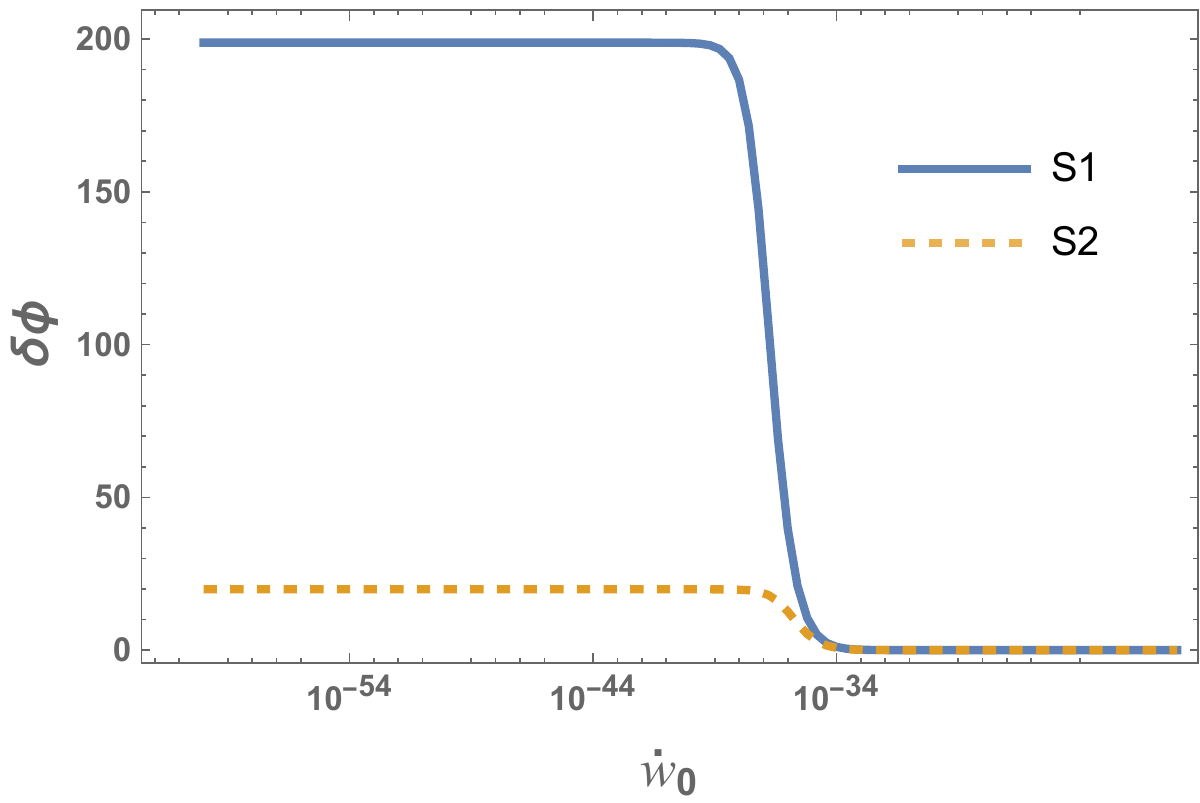}\label{f13a}}
\hspace{10mm}
\subfloat[The initial geodesic deviation velocity is negative.]{\includegraphics[scale=0.39]{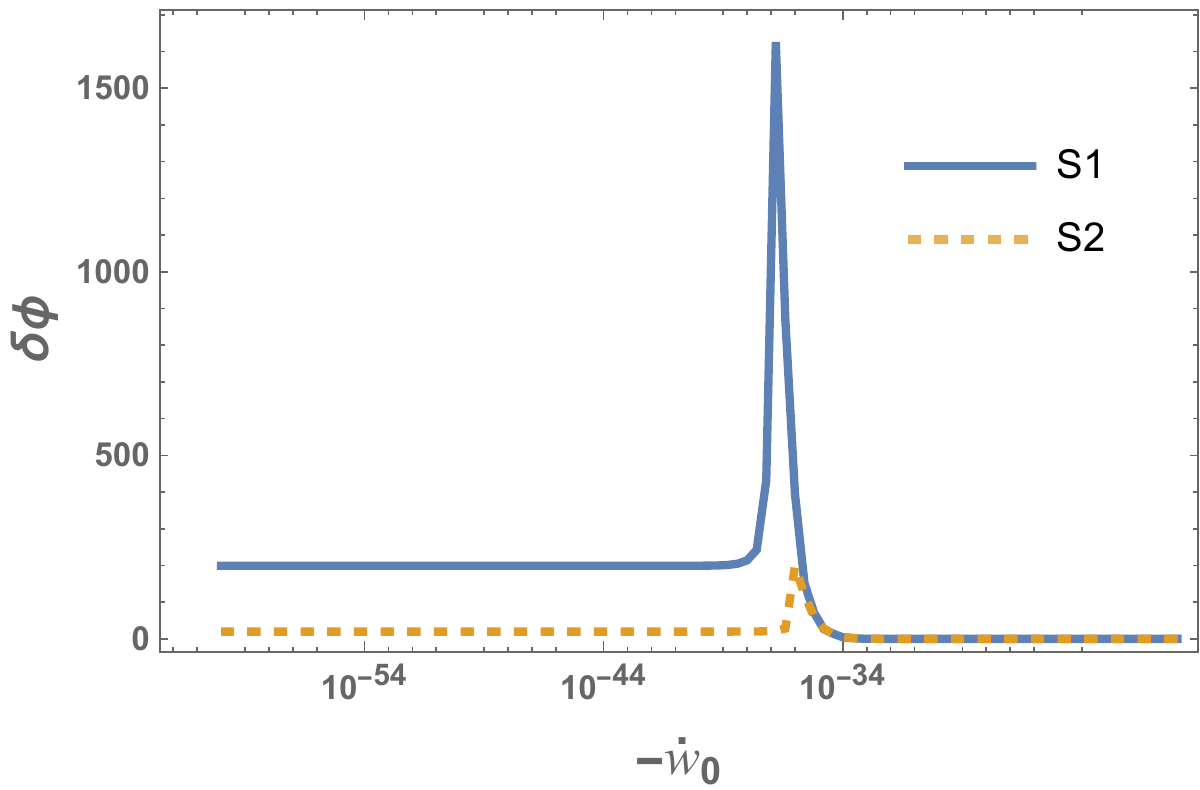}\label{f13b}}
\caption{The phase change $\delta \phi$ as a function of initial geodesic deviation velocity $\dot{w}_0$ of the particles. The initial values of the other parameters are given in Table \ref{initial}.}\label{f13}
\end{figure*}

As we can see from Fig. \ref{f19}, the phase change is just inversely proportional to $w_0$ as expected.

Fig. \ref{f13} is a plot of $\delta \phi$ as a function of initial geodesic deviation velocity $\dot{w}_0$ of the particles\footnote{Again, for simplicity, we have assumed that $\dot{w}_0^{(1)}=\dot{w}_0^{(2)}=\dot{w}_0$. }. The initial values of other parameters are given in Table \ref{initial}. From Fig. \ref{f13a}, we find that for smaller $\dot{w}_0$, $\delta\phi$ decreases slowly with increasing $\dot{w}_0$, then suddenly for some point ($\dot{w}_0 \approx {10^{ - 37}}$), it decrease quickly and then tends to be nearly flat. This behavior can be roughly explained in the following way:  Eq. \eqref{14} can be converted to $\frac{{{d^2}{w^{(a)} }}}{{d{\tau ^2}}}{\rm{ =  - }}{k^{(a)} }_{{(b)}} {w^{(b)} }$, the r.h.s of the equation looks like a ``geodesic deviation acceleration'' which is proportional to ${k^{(a)} }_{(b)}$ and ${w^{(b)} }$. Meanwhile, comparing with other regions, ${k^{(a)} }_{(b)}$ has dominant effects near the center of the cluster. As a consequence, the particle pair with a larger ${w^{(b)} }$ will acquire an overwhelming geodesic deviation acceleration as they pass the center of the cluster. The separation distance $d(\tau)$ between them will continue to increase after this process. So when the absolute value of the geodesic deviation velocity reaches a certain value, ${w^{(b)} }$ (or $d(\tau)$) will be large enough and the phase change will drop sharply as it is inverse proportional to $d(\tau)$ as shown in Eqs. \eqref{17} and \eqref{18}.

The negative initial geodesic deviation velocity case, however, is very different as shown in Fig. \ref{f13b}. $\delta\phi$ of this case increases slowly with increasing $-\dot{w}_0$ at the beginning,  and then suddenly increases to a maximum at some point ($\dot{w}_0 \approx {10^{ - 37}}$), followed by a quick decrease to lower value. Finally it descends gradually to some finite value. Recalling a similar behavior existing in S3 of Fig. \ref{5a}, where $d(\tau)$ firstly goes down quickly as increasing  $-\dot{w}_0$ and then grows up vastly. We know that this is because for the particle pairs with a negative initial geodesic deviation velocity, at the beginning the particle pairs approach each other and then they move away from each other. Similarly, in the present case, the particles with a negative initial geodesic deviation velocity approach each other firstly and later will move away from each other, so that $d(\tau)$ will decrease quickly at the beginning and then increase fast, leading to inverse behavior of $\delta\phi$ as $d(\tau)$ appearing in the denominator of Eq. \eqref{17}.

\begin{figure}\label{f16}\centering
\subfloat[]{\includegraphics[scale=0.5]{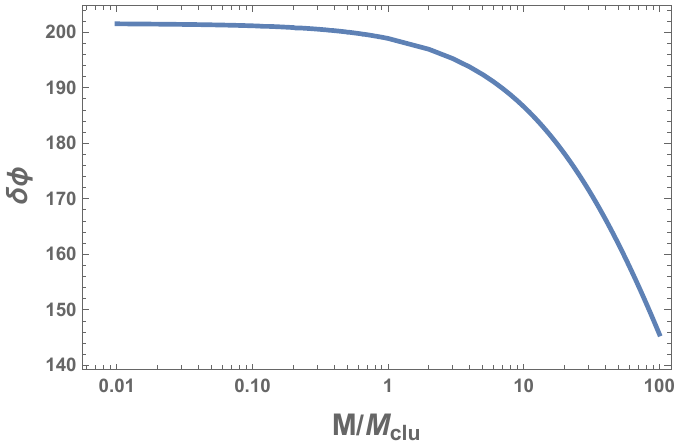}\label{f16a}}\\
\subfloat[]{\includegraphics[scale=0.5]{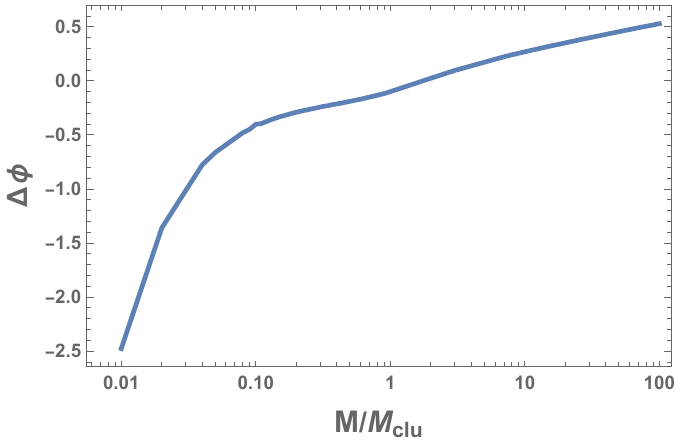}\label{f16b}}
\caption{(a) The phase change $\delta \phi$ as a function of mass for dPIE model. (b) The difference of the phase change for different models $\Delta \phi=\delta\phi_{dPIE}-\delta\phi_{NFW}$ as a function of $M$. Here, $M_{clu}$ is just the mass given in Table \ref{table1}. The initial values of the other parameters are given by S1 in Table \ref{initial}.} 
\end{figure}

In the previous simulations, we let the total mass $M$ of the cluster fix. In order to see the influence of $M$ on the entanglement phase, it is useful to calculate the phase change with different mass in dPIE and NFW model. We first plot phase change as a function of $M$ in the dPIE model as shown in Fig.  \ref{f16a}. Then we define a $\Delta \phi=\delta\phi_{dPIE}-\delta\phi_{NFW}$, where $\delta\phi_{dPIE}$ and $\delta\phi_{NFW}$ denote, respectively, the phase change for dPIE and NFW models. We then plot $\Delta \phi$ as a function of $M$ as shown in Fig.  \ref{f16b}.
We can find from Fig.  \ref{f16a} that phase change decreases faster and faster as $M$ increases. This is because the stronger the gravitational field, the shorter the proper time, the phase change decreases with galaxy's mass. Then we can see from Fig.  \ref{f16b} that the specific model also act on the phase change in a certain way. In the smaller mass range NFW model is more conducive to the formation of phase change while in the bigger mass range dPIE is better at inducing phase change.

In summary, compared with dPIE model, the relationship between the phase change and each initial value exhibits the similar characteristics. Generally speaking, larger $z$, smaller initial radial velocity $v_0$, and appropriate negative initial deviation velocity lead to bigger $\delta\phi$ and (possibly) more significant effects of entanglement.

\section{Offset geodesics and characteristic spectrum}
In the preceding discussion, we made the assumption that all particles follow geodesics passing through the cluster's center. However, in reality, geodesics may deviate from the cluster's center in various ways. Moreover, from an experimental standpoint, distinguishing whether observed entanglement arises from the quantum gravity effect of particle pairs or from their emission at the source presents a significant challenge. Therefore, it becomes crucial to discern whether gravity induces entanglement or if other physical processes are at play.

To tackle this challenge, we propose exploring the characteristic spectrum of QGEM during geodesic motion, which manifests as phase changes along a series of geodesics. Through analysis of the entangled patterns formed by different geodesics, we can infer the entanglement's origin—whether it arises from the gravitational field of nearby particles or from alternative sources. 

\begin{figure*}\centering
\subfloat[]{\includegraphics[scale=0.4]{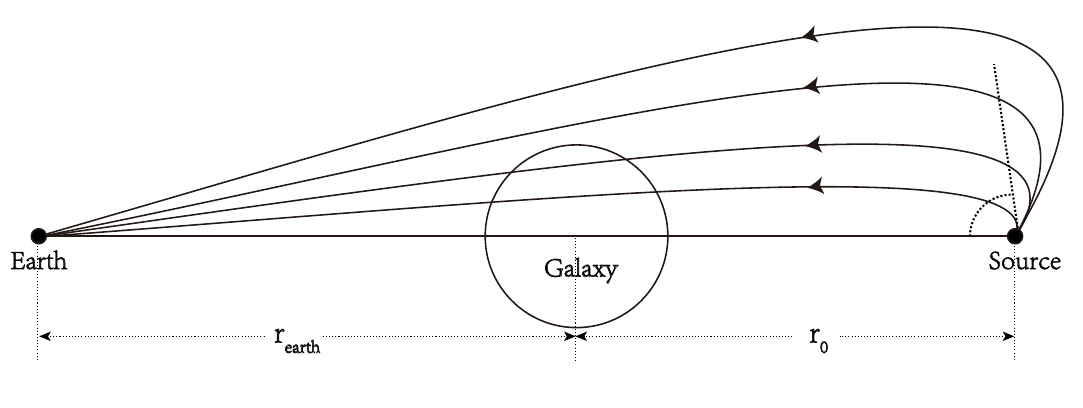}}
\subfloat[]{\includegraphics[scale=0.4]{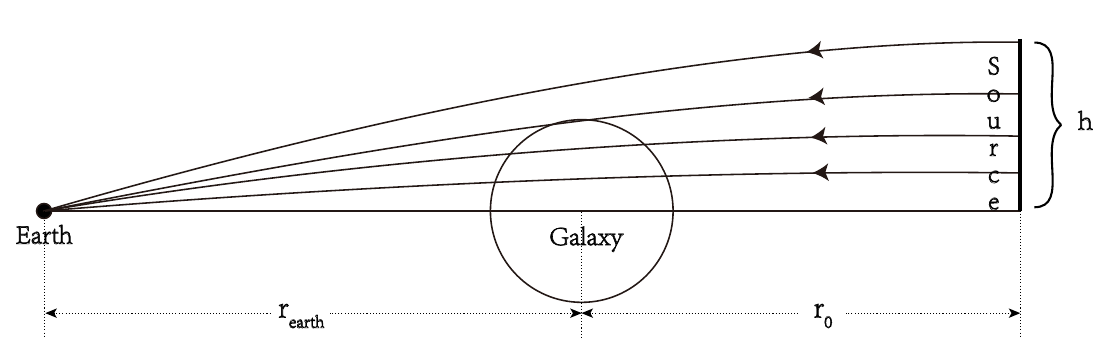}\label{offset2}}
\caption{Two series geodesics offset from the center of the galaxy. (a) The particles' source location is given. (b) The particles' source location is unknown.}\label{f4}
\end{figure*}

Generally speaking, there are two possible geodesics, contingent upon our knowledge of the particles' source location, as illustrated in Fig. \ref{f4}. In what follows, we would like to discuss possible characteristic spectrum of the QGEM in these two cases, respectively.

\subsection{The particles' source location is given}

In the first geodesic case, we consider a globular cluster with a density profile described by the dPIE model \cite{r8}. The parameter values of the model are chosen the same as in Table \ref{table1}, and the particle's mass is assumed, again, to be \( {10^{ - 25}}\text{kg}\). For the sake of simplicity, in this subsection, we only consider the set S1 from Table \ref{initial} as the initial parameter values. 

At the source point, we use the same orthogonal frame as in \eqref{8} and \eqref{8b}. The four-velocity is expressed as:
\beq\label{4v}
{u^{\mu}} = \gamma {e_{(0)}}^{\mu} - \gamma v_0 \left( {\cos \xi {e_{(1)}}^{\mu} - \sin \xi {e_{(2)}}^{\mu}} \right),
\eeq
where $\gamma = \frac{1}{{\sqrt {1 - v_0^2} }}$  and $v_0$
 is the particle's initial emission velocity.  The four-velocity mentioned above is connected to the deflection angle $\xi$, with $\xi$ being assigned to values in the range of $\xi \in [0, \pi)$ for the initial four-velocity. For particles that can reach the Earth, $\xi$ and $v_0$ are not independent. Namely, for each $\xi$, there is a unique value of $v_0$. To simplify our analysis, we will focus on the scenario where the entire trajectories lie in the same plane. Under these initial conditions, the geodesics will deviate from the center of the cluster. Notably, when $\xi = 0$, the geodesics that pass through the galaxy's center will be recovered. We also build a set of parallelly transported tetrads and assume the initial geodesic deviation vector is:
 \beq\label{52}
 {w^{(a)}} = {{\tilde e}_{(\parallel )}}{w^\parallel } + {{\tilde e}_{( \bot )}}{w^ \bot },
 \eeq
where ${{\tilde e}_{(\parallel )}}$ is the basis parallel to the connecting line between the Earth and the galaxy cluster , ${{\tilde e}_{( \bot )}}$ is the basis perpendicular to ${{\tilde e}_{(\parallel )}}$ and four-velocity. The initial geodesic deviation component ${w^\parallel }= \frac13\times 10^{-16}$ and ${w^ \bot } = 0$.
 
 Now let us delve into the QGEM effect in this scenario. As usual, we can denote positive helicity as $\uparrow$ and negative helicity as $\downarrow$. Therefore,  the adjacent propagating particles are in a superposition state of $\frac{1}{{\sqrt 2 }}\left( {\left| { \uparrow L} \right\rangle + \left| { \downarrow R} \right\rangle } \right)$. As a result, the expected entangled final state matches \eqref{fstate}, and the phase change induced by gravity is again described by \eqref{dphi}. Following the same procedures given in section III, one can  calculate the phase change under different initial conditions. 

Compared to the initial conditions, particle's energy is an important observable. In observer's frame, each particle emitted at a different angle will have a specific speed, and the Earth's observer will measure a corresponding kinetic energy. In this static spherically symmetric space-time, there is a conserved quantity along geodesics, denoted as $C$ \cite{r4}, is given by \footnote{This is because: for a given Killing vector $\xi^{\mu}$ and a tangent vector $u^{\mu}=(\partial_{\tau})^{\mu}$ of a geodesic $\gamma(\tau)$, the identity holds: $u^{\mu}\nabla_{\mu}(\xi_{\nu}u^{\nu})=0$.}:
\beq\label{5}
C = {\xi _{\mu}}{u^{\mu}} = \frac{{\sqrt {{\xi_{\mu}}{\xi ^{\mu}}} }}{m}{E_{ob}},
\eeq
where ${{\xi ^{\mu}}}$ represents the timelike killing vector $(\partial t )^{\mu}$, ${u^{\mu}}$ is the particle's four-velocity \eqref{4v}, and $E_{ob}$ is the energy measured by a local stationary observer at infinity (e.g., an observer on Earth). We use the kinetic energy of the particles measured by observer on Earth to label each particle. The observed kinetic energy for particles emitted from different initial angles is calculated as follows:
\bqn\label{6}
{E_{\text{kin}}} &=& E_{ob}-m \nonumber\\
&=& m\gamma e^{A(r_0)-A(r_{\text{earth}})}- m,
\eqn
where, ${r_{\text{earth}}}$ represents the coordinate distance between the center of the Earth and the center of the galaxy cluster, and ${r_{0}}$ represents the coordinate distance between location of particle emission and the center of the galaxy cluster.

\begin{figure}\centering
\includegraphics[scale=0.4]{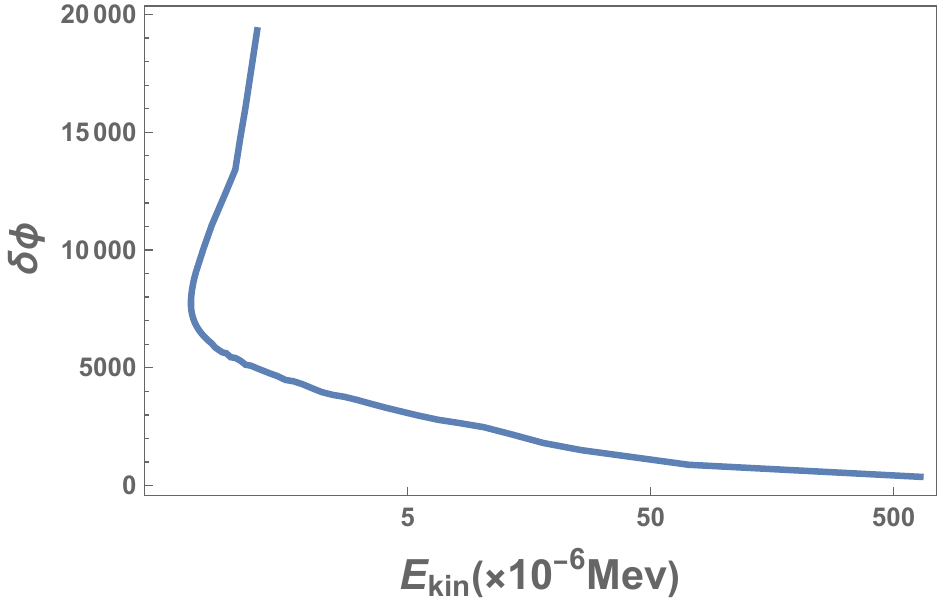}
\caption{Plot of entangled phase and observed particle energy. The vertical axis represents the entanglement phase $\delta \phi$, which is formed in the propagation process. On the horizontal axis, we have the kinetic energy $E_{\text{kin}}$ measured by Earth observers after the particles with different emission parameters have reached the Earth. }\label{phase_energy1}
\end{figure}

The result is shown in Fig.\ref{phase_energy1}. It shows that as the emission angle $\xi$ increases, the kinetic energy of particles reaching Earth decreases, and the intrinsic length of the geodesic line increases, leading to a progressively larger entanglement phase. The irregular jumps on the curve are caused by the instability of numerical simulation. 

\begin{figure}\centering
\includegraphics[scale=0.4]{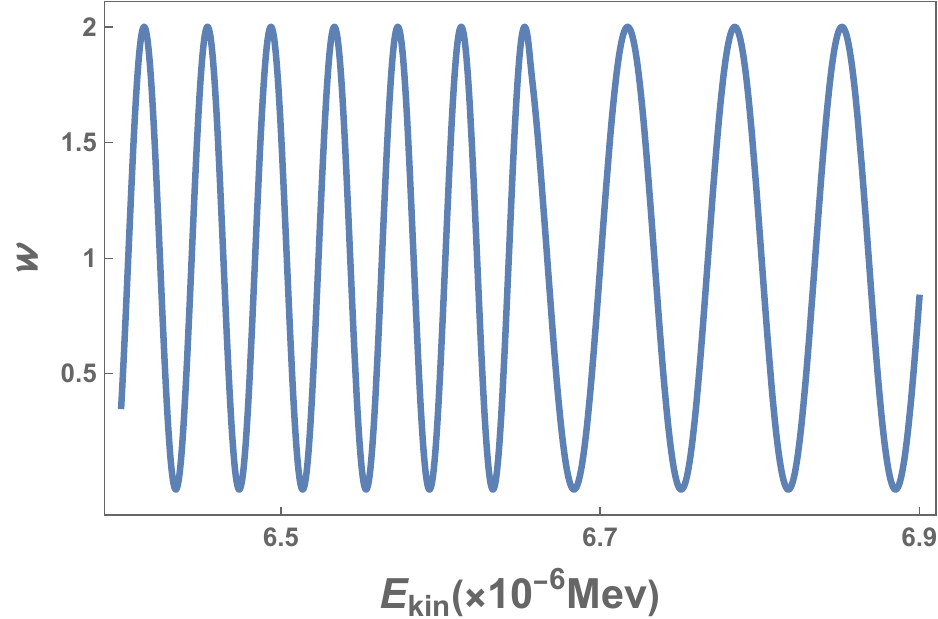}
\caption{Variation of entanglement witness $\mathcal{W}$ as a function of $E_{kin}$ in dPIE model.}\label{wvse}
\end{figure}

Phase change is not directly observable. On the other hand, the entanglement witness $\mathcal{W}$ is often utilized as an experimental indicator to detect entanglement formation.The definition of entanglement witness is
\beq\label{51}
 \mathcal{W} = \left| {\left\langle {\sigma {1_x} \otimes \sigma {2_z}} \right\rangle  + \left\langle {\sigma {1_y} \otimes \sigma {2_y}} \right\rangle } \right|.
\eeq
When it's greater than 1, we can infer that there is entanglement between the two particles. The variation of $\mathcal{W}$ with respect to $E_{kin}$ is plotted in Fig. \ref{wvse}. From this figure, we observe that the entanglement witness oscillates rapidly with energy, but it exhibits different oscillating frequencies in various energy segments. Specifically, it oscillates relatively fast in the region of low energy, which coincides with the fact that the phase increases faster with an increase in the emission angle.

\subsection{The particles' source location is unknown}
In this case, since we lack information about the location of the particles' source, we cannot use the deflection angle to span the geodesics. Instead, we utilize $h$, which represents the initial vertical distance from horizontal geodesics, to effectively span the geodesics, as illustrated in Fig. \ref{offset2}. Each particle exhibits a distinct starting $h$ value and, when coupled with an appropriate initial four-velocity, can be successfully detected by a probe on Earth. The four-velocity can be expressed as follows:
\beq\label{9}
{u^{\mu}} = \gamma {e_{(0)}}^{\mu} + \gamma v_0 {e_\parallel }^{\mu},
\eeq
where ${e_\parallel}^{\mu}$ is the unit space-like base formed by the linear combination of ${e_{(1)}}^{\mu}$ and ${e_{(2)}}^{\mu}$, making it parallel to the connecting line between the Earth and the center of the galaxy cluster. Namely,
\bqn\label{50}
{e_\parallel }^\mu  &=& \sqrt {\frac{{{r_0}^2 + {h^2}}}{{{e^{2B\left( {{r_0}^2 + {h^2}} \right)}}{r_0}^2 + {h^2}}}}\nonumber\\
&&\cdot \left( { - \frac{{{r_0}}}{{\sqrt {{r_0}^2 + {h^2}} }}{e_{(1)}}^\mu  + \frac{h}{{{r_0}^2 + {h^2}}}{e_{(2)}}^\mu } \right)
\eqn

In this case, the kinetic energy observed by the observer on Earth is determined by 
\beq\label{90}
{E_{{\rm{kin}}}}=m\gamma {e^{A(\sqrt {{r_0}^2 + {h^2}} ) - A({r_{{\rm{earth}}}})}} - m ,
\eeq
and each geodesic to the Earth corresponds to a specific value of $ v_0 $ and $ \gamma $.

\begin{figure}\centering
\includegraphics[scale=0.4]{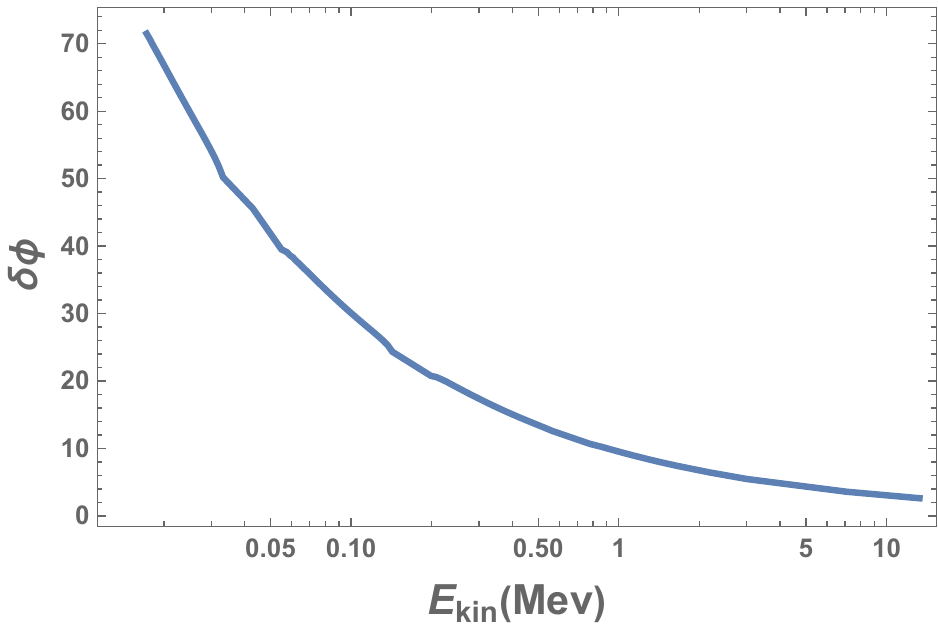}
\caption{Entangled phase as a function of observed particle energy. The vertical axis represents the entanglement phase $\delta \phi$ formed during the propagation process, while the horizontal axis corresponds to the kinetic energy $E_{kin}$ measured by the Earth observer after the particles with different emission parameters reach the Earth.}\label{phik}
\end{figure}

\begin{figure}\centering
\includegraphics[scale=0.4]{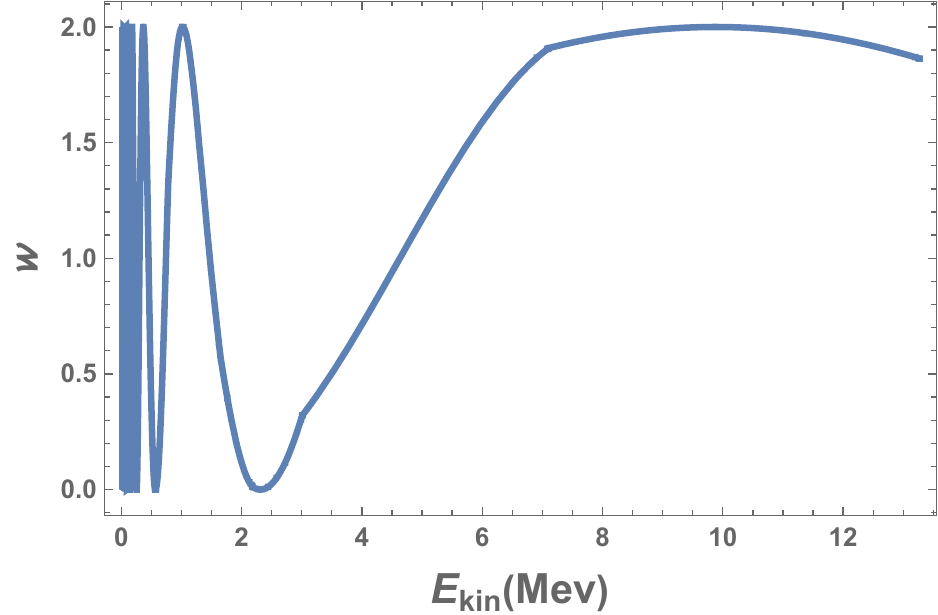}
\caption{Entanglement witness as a function of kinetic energy.}\label{wvk}
\end{figure}

\begin{figure}\centering
	\includegraphics[scale=0.4]{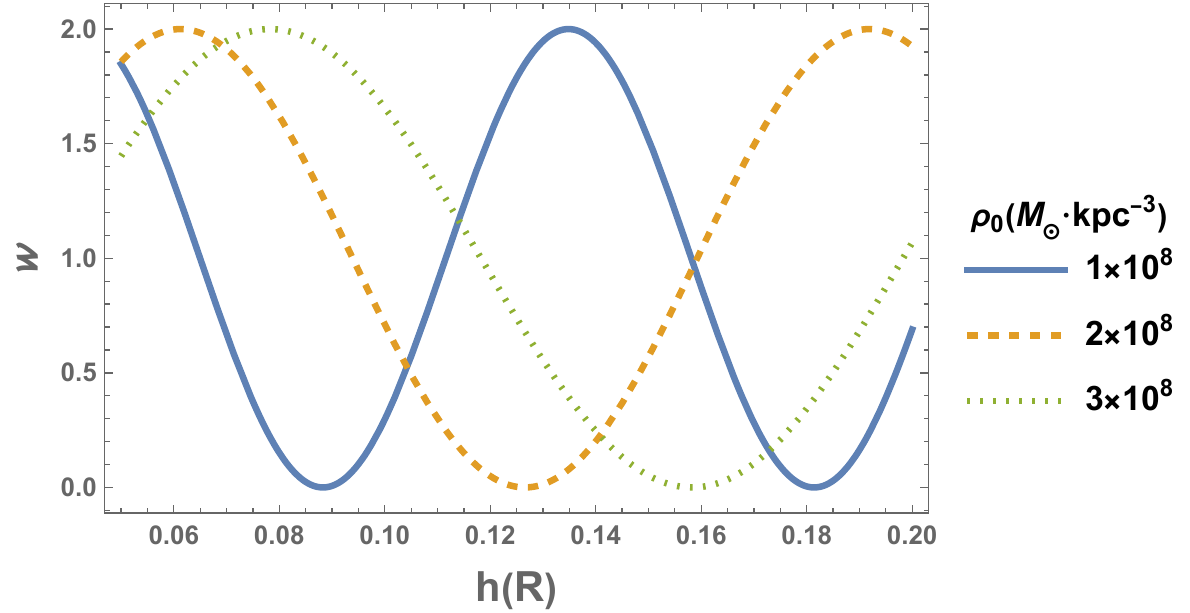}
	\caption{Variation of entanglement witness with particle emission height for dPIE models with different center density parameters. Here, we assume that the value range of $h$ is 0.05$R$ to 0.2$R$. Other galaxy parameters and kinematic parameters remain unchanged.}\label{wvh}
\end{figure}

The characteristic entangled spectral lines are shown in Figure \ref{phik} and Fig. \ref{wvk}. Similarly, the entanglement phase drops rapidly with increasing energy, and the entanglement witness oscillates slowly at high energy and rapidly at low energy. This behavior indicates that at greater emission heights, the phase increases faster. The plot of entanglement witness as a function of the emission height $h$ can show more details. From Fig. \ref{wvh}, we see that for dPIE models with different center densities, the segment of the entanglement curve where the entanglement exceeds $1$ varies. Additionally, the entanglement witness with greater center density oscillates more slowly. In other words, a higher center density ${\rho_0}$ of a galaxy will impede the generation of entanglement between particles. A possible explanation is that, under the condition of the same emission height, particles take less proper time to reach Earth when passing through more massive galaxies, which hinders entanglement formation.

In summary, in both cases, the entanglement phase varies monotonously with the initial geodesic emission parameters, $\xi$ and $h$, causing the entanglement witness $\mathcal{W}$ to oscillate with the initial parameters (or the kinetic energy), as shown in Figs. \ref{wvse}, \ref{wvk} and \ref{wvh}. This characteristic could be a significant index to distinguish the origin of the observed entanglement. More specifically, if the particles' entanglement is native (formed during particle emission), then the entanglement witness will be randomly distributed with energy and will not exhibit the quasi-periodic oscillation behavior mentioned above. In other words, only entanglement induced by quantum gravity effects can result in $\mathcal{W}$ exhibiting this quasi-periodic behavior. Thus, the identification of these entanglement features in experimental observations can provide confidence in concluding that the entanglement is induced by quantum gravity along the geodesic, rather than being formed during their emission.

\begin{figure}\centering
\includegraphics[scale=0.4]{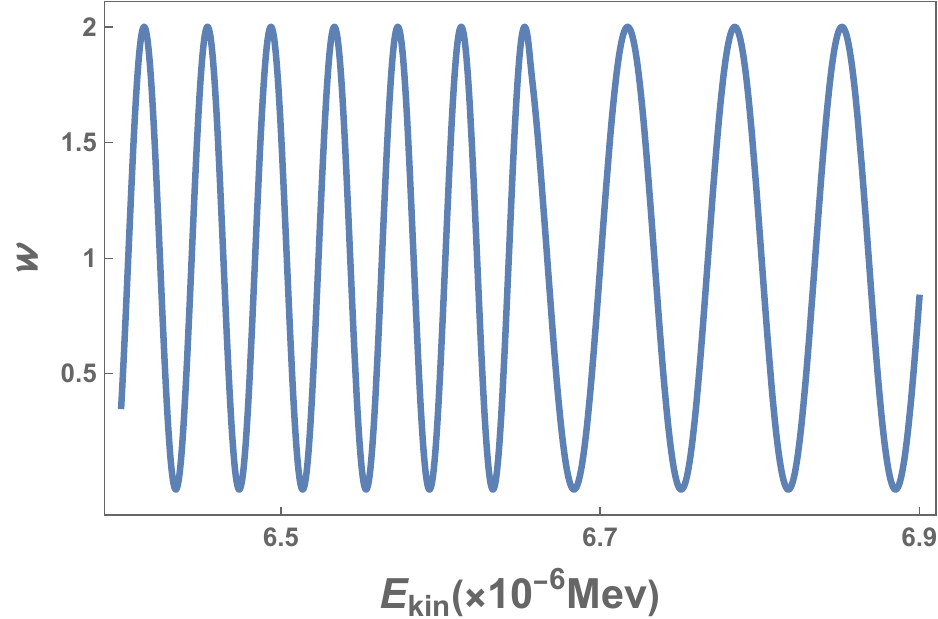}
\caption{Entanglement witness as a function of kinetic energy in the NFW model as the location of the particles' source is given.}\label{ew_nfw1}
\end{figure}

\begin{figure}\centering
\includegraphics[scale=0.4]{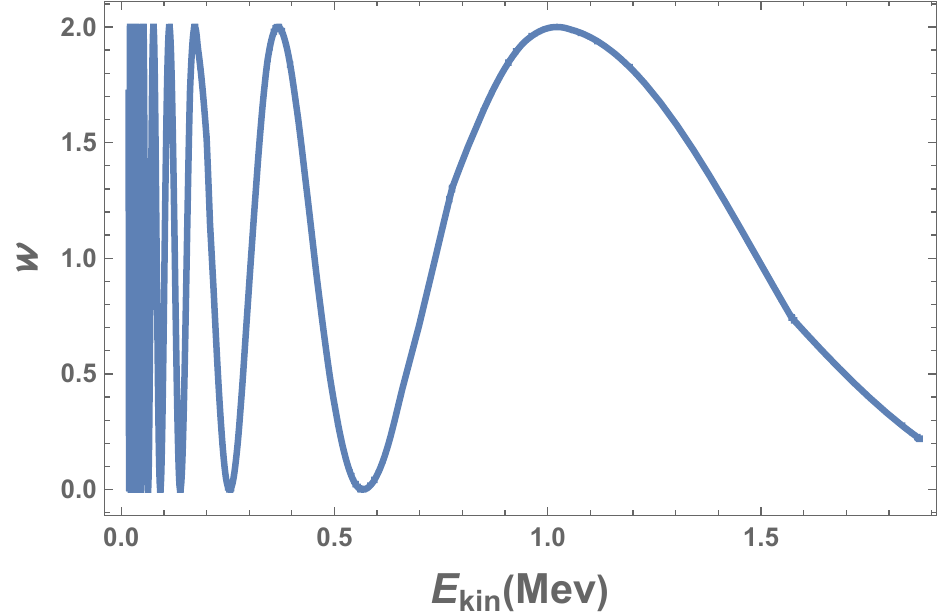}
\caption{Entanglement witness as a function of kinetic energy in the NFW model as the location of the particles' source is not given.}\label{ew_nfw2}
\end{figure}

The above conclusions are also valid for the NFW model. Fig. \ref{ew_nfw1} and Fig. \ref{ew_nfw2} show the entanglement witness as a function of kinetic energy, using the initial NFW parameters given in Tables \ref{table1} and \ref{initial}. These plots explicitly exhibit similar quasi-oscillation behavior as observed in the dPIE model in BCG.

\section{Conclusions and discussions}
In the realm of quantum gravity, the challenge lies not in the absence of a complete mathematical physical theory, but rather in the scarcity of experimental approaches to connect theory and reality. In light of recent advancements in quantum technologies, endeavors have been made to elucidate the quantum nature of gravity. Among these endeavors, the quantum gravity induced entanglement of masses (QGEM) proposal has garnered significant attention \cite{r2,r3,r4,r6}. In this paper, we extend the QGEM experiments to include curved spacetimes. More specifically, we consider the QGEM for a pair of particles traveling along their geodesics in a galaxy with Schwarzschild metric as the spacetime background. We find that particle pairs readily become entangled at larger radial coordinates with appropriate small initial radial velocities.

By investigating the relations between entanglement witness and the kinetic energy observed by the observer (determined by the initial values of the model parameters), we find that there has a characteristic spectrum of QGEM. This provides us a way to distinguish whether the observed entanglement arises from the quantum gravity effect of particle pairs or from other process (e.g, the emission stage at the source as suggested by the Hawking radiation of black holes).

In addition, execution of the proposals outlined in this paper will demonstrate a more comprehensive quantum gravity effect in extensive spacetime, surpassing the limitations of local experiments carried out in labs \cite{r2,r3}. Notably, our scheme accommodates much lighter particle masses and significantly larger particle spacings. These particles may traverse hundreds of millions of years of geodesic movement before detection, an impractical feat on Earth. Additionally, direct detection of particles from the universe simplifies the experiment's preparation process. Microscopic particles may spontaneously exist in a superposition state somewhere in the universe, coinciding with our computational hypothesis mentioned above.

Furthermore, our astronomical scheme, based on the hypothesis of the equivalence principle in superposition spacetime \cite{Giacomini:2020ahk}, hopefully serves as an alternative validation for this extended equivalence principle.

Admittedly, there are some technical details need improve in the future. The geodesic deviation equation serves as a first-order approximation formula for describing spacelike distance variations between particles. We solely consider the background spacetime metric while neglecting the backreactions of the nearby particles. Numerical calculations introduce method error and truncation error in the numerical simulation. Additionally, our idealized assumptions may not fully capture the complexities of the real universe. 

In future research, we aim to employ more sophisticated calculation methods to account for the gravitational force of nearby particles. Additionally, we will explore more complex geodesic trajectories beyond radial movement to detect as many entangled particles as possible. Furthermore, in cosmic spacetime, two particles may be separated by considerable spacelike distances. As highlighted in \cite{r6}, we will seek new methods to extract spacelike entanglement and investigate how gravity induces spacelike entanglement in curved spacetime.

 \section*{Acknowledgements}
This work was supported by the National Natural Science Foundation of China with the Grants No. 12375049, 11975116, and Key Program of the Natural Science Foundation of Jiangxi Province under Grant No. 20232ACB201008. 
\appendix
\section{Inner metric of BCG}\label{inner metric}
Inside a static globular galaxy cluster, the metric could be assumed as:
\begin{equation}\label{A4}
d{s^2} = {e^{2A(r)}}d{t^2} - {e^{2B(r)}}d{r^2} - {r^2}\left( {d{\theta ^2} + {{\sin }^2}\theta {d^2}\varphi } \right),
\end{equation}
where ${A(r)}$ is calculated by $\mathop {A(r) = \smallint }\nolimits_0^r \frac{{m\left( x \right)}}{{{x^2}}}{\rm{d}}x + c$. The additional $c$ is added so that ${A(r)}$ transitions smoothly in and out of the galaxy. Substituting \eqref{2} into above formula, after careful calculation, we find the full expression for $A(r)$ is
 \begin{widetext}
\begin{equation}\label{A5}
A(r) = \frac{{2\pi {\rho _0}r_{\text{core}}^2r_{\text{cut}}^2\left( {r\left( {\log \left( {\frac{{{r^2}}}{{r_{\text{cut}}^2}} + 1} \right) - \log \left( {\frac{{{r^2}}}{{r_{\text{core}}^2}} + 1} \right)} \right) - 2{r_{\text{core}}}\arccot\left( {\frac{{{r_{\text{core}}}}}{r}} \right) + 2{r_{\text{cut}}}\arccot\left( {\frac{{{r_{\text{cut}}}}}{r}} \right)} \right)}}{{r\left( {r_{\text{core}}^2 - r_{\text{cut}}^2} \right)}} + c,
\end{equation}
 \end{widetext}
 \begin{widetext}
 where
\begin{equation}\label{A6}
\begin{split}c =& \frac{{2\pi {\rho _0}r_{\text{core}}^2r_{\text{cut}}^2\left( {R\left( {\log \left( {\frac{{{R^2}}}{{r_{\text{core}}^2}} + 1} \right) - \log \left( {\frac{{{R^2}}}{{r_{\text{cut}}^2}} + 1} \right)} \right) + 2{r_{\text{core}}}{{\cot }^{ - 1}}\left( {\frac{{{r_{\text{core}}}}}{R}} \right) - 2{r_{\text{cut}}}{{\cot }^{ - 1}}\left( {\frac{{{r_{\text{cut}}}}}{R}} \right)} \right)}}{{R\left( {r_{\text{core}}^2 - r_{\text{cut}}^2} \right)}} \\&+ 0.5\log \left( {\frac{{8\pi {\rho _0}r_{\text{core}}^2r_{\text{cut}}^2\left( {{r_{\text{cut}}}{{\tan }^{ - 1}}\left( {\frac{R}{{{r_{\text{cut}}}}}} \right) - {r_{\text{core}}}{{\tan }^{ - 1}}\left( {\frac{R}{{{r_{\text{core}}}}}} \right)} \right)}}{{R\left( {r_{\text{core}}^2 - r_{\text{cut}}^2} \right)}} + 1} \right).\end{split}
\end{equation}
 \end{widetext}


\begin{thebibliography}{99}
\bibitem{r1}S. Hossenfelder, Experimental Search for Quantum Gravity, Springer, 2017.

\bibitem{r2}
S.~Bose, A.~Mazumdar, G.~W.~Morley, H.~Ulbricht, M.~Toro\v{s}, M.~Paternostro, A.~A.~Geraci, P.~F.~Barker, M.~S.~Kim and G.~Milburn,
``Spin Entanglement Witness for Quantum Gravity,''
Phys. Rev. Lett. \textbf{119}, no.24, 240401 (2017)
[arXiv:1707.06050 [quant-ph]].
\bibitem{r3}
C.~Marletto and V.~Vedral,
``Gravitationally induced entanglement between two massive particles is sufficient evidence of quantum effects in gravity,''
Phys. Rev. Lett. \textbf{119}, no.24, 240402 (2017)
[arXiv:1707.06036 [quant-ph]].

\bibitem{r4}
M.~Christodoulou and C.~Rovelli,
``On the possibility of laboratory evidence for quantum superposition of geometries,''
Phys. Lett. B \textbf{792}, 64-68 (2019)
[arXiv:1808.05842 [gr-qc]].

\bibitem{Capolupo:2019peg} 
A.~Capolupo, G.~Lambiase, A.~Quaranta and S.~M.~Giampaolo,
``Probing axion mediated fermion\textendash{}fermion interaction by means of entanglement,''
Phys. Lett. B \textbf{804}, 135407 (2020)
[arXiv:1910.01533 [hep-ph]].

\bibitem{Belenchia:2018szb}
A.~Belenchia, R.~M.~Wald, F.~Giacomini, E.~Castro-Ruiz, \v{C}.~Brukner and M.~Aspelmeyer,
``Quantum Superposition of Massive Objects and the Quantization of Gravity,''
Phys. Rev. D \textbf{98}, no.12, 126009 (2018)
[arXiv:1807.07015 [quant-ph]].
\bibitem{Carney:2018ofe}
D.~Carney, P.~C.~E.~Stamp and J.~M.~Taylor,
``Tabletop experiments for quantum gravity: a user\textquoteright{}s manual,''
Class. Quant. Grav. \textbf{36}, no.3, 034001 (2019)
[arXiv:1807.11494 [quant-ph]].
\bibitem{Marshman:2019sne}
R.~J.~Marshman, A.~Mazumdar and S.~Bose,
``Locality and entanglement in table-top testing of the quantum nature of linearized gravity,''
Phys. Rev. A \textbf{101}, no.5, 052110 (2020)
[arXiv:1907.01568 [quant-ph]].
\bibitem{Buoninfante:2018xiw}
L.~Buoninfante, A.~S.~Koshelev, G.~Lambiase and A.~Mazumdar,
``Classical properties of non-local, ghost- and singularity-free gravity,''
JCAP \textbf{09}, 034 (2018)
[arXiv:1802.00399 [gr-qc]].
\bibitem{Westphal:2020okx}
T.~Westphal, H.~Hepach, J.~Pfaff and M.~Aspelmeyer,
``Measurement of gravitational coupling between millimetre-sized masses,''
Nature \textbf{591}, no.7849, 225-228 (2021)
[arXiv:2009.09546 [gr-qc]].
\bibitem{Carlesso:2019cuh}
M.~Carlesso, A.~Bassi, M.~Paternostro and H.~Ulbricht,
``Testing the gravitational field generated by a quantum superposition,''
New J. Phys. \textbf{21}, no.9, 093052 (2019)
[arXiv:1906.04513 [quant-ph]].
\bibitem{Danielson:2021egj}
D.~L.~Danielson, G.~Satishchandran and R.~M.~Wald,
``Gravitationally mediated entanglement: Newtonian field versus gravitons,''
Phys. Rev. D \textbf{105}, no.8, 086001 (2022)
[arXiv:2112.10798 [quant-ph]].
\bibitem{Christodoulou:2022mkf}
M.~Christodoulou, A.~Di Biagio, M.~Aspelmeyer, \v{C}.~Brukner, C.~Rovelli and R.~Howl,
``Locally Mediated Entanglement in Linearized Quantum Gravity,''
Phys. Rev. Lett. \textbf{130}, no.10, 100202 (2023)
[arXiv:2202.03368 [quant-ph]].
\bibitem{Cho:2021gvg}
H.~T.~Cho and B.~L.~Hu,
``Quantum noise of gravitons and stochastic force on geodesic separation,''
Phys. Rev. D \textbf{105}, no.8, 086004 (2022)
[arXiv:2112.08174 [gr-qc]].

\bibitem{Matsumura:2020law}
A.~Matsumura and K.~Yamamoto,
``Gravity-induced entanglement in optomechanical systems,''
Phys. Rev. D \textbf{102}, no.10, 106021 (2020)
[arXiv:2010.05161 [gr-qc]].

\bibitem{Miki:2020hvg}
D.~Miki, A.~Matsumura and K.~Yamamoto,
``Entanglement and decoherence of massive particles due to gravity,''
Phys. Rev. D \textbf{103}, no.2, 026017 (2021)
[arXiv:2010.05159 [gr-qc]].

\bibitem{Howl:2023xtf}
R.~Howl, N.~Cooper and L.~Hackerm\"uller,
``Gravitationally-induced entanglement in cold atoms,''
[arXiv:2304.00734 [quant-ph]].

\bibitem{Yant:2023smr}
J.~Yant and M.~Blencowe,
``Gravitationally induced entanglement in a harmonic trap,''
Phys. Rev. D \textbf{107}, no.10, 106018 (2023)
[arXiv:2302.05463 [quant-ph]].

\bibitem{Feng:2023krm}
S.~Feng, B.~M.~Gu and F.~W.~Shu,
``Detecting Extra Dimension By the Experiment of the Quantum Gravity Induced Entanglement of Masses,''
[arXiv:2307.11391 [gr-qc]].

\bibitem{Schut:2023eux}
M.~Schut, A.~Grinin, A.~Dana, S.~Bose, A.~Geraci and A.~Mazumdar,
``Relaxation of experimental parameters in a Quantum-Gravity Induced Entanglement of Masses Protocol using electromagnetic screening,''
[arXiv:2307.07536 [quant-ph]].

\bibitem{Fragolino:2023agd}
P.~Fragolino, M.~Schut, M.~Toro\v{s}, S.~Bose and A.~Mazumdar,
``Decoherence of a matter-wave interferometer due to dipole-dipole interactions,''
[arXiv:2307.07001 [quant-ph]].

\bibitem{Li:2022yiy}
P.~Li, Y.~Ling and Z.~Yu,
``Generation rate of quantum gravity induced entanglement with multiple massive particles,''
Phys. Rev. D \textbf{107}, no.6, 064054 (2023)
[arXiv:2210.17259 [gr-qc]].

\bibitem{Bose:2022uxe}
S.~Bose, A.~Mazumdar, M.~Schut and M.~Toro\v{s},
``Mechanism for the quantum natured gravitons to entangle masses,''
Phys. Rev. D \textbf{105}, no.10, 106028 (2022)
[arXiv:2201.03583 [gr-qc]].

\bibitem{He:2023hys}
F.~He and B.~Zhang,
``Generation of entanglement between two laser pulses through gravitational interaction,''
Eur. Phys. J. Plus \textbf{138}, no.2, 141 (2023)
[arXiv:2302.06362 [gr-qc]].

\bibitem{Giacomini:2020ahk}
F.~Giacomini and \v{C}.~Brukner,
``Einstein's Equivalence principle for superpositions of gravitational fields and quantum reference frames,''
[arXiv:2012.13754 [quant-ph]].

\bibitem{Joos:1984uk}
E.~Joos and H.~D.~Zeh,
Z. Phys. B \textbf{59}, 223-243 (1985)
doi:10.1007/BF01725541

\bibitem{Arrasmith:2017ogi}
A.~Arrasmith, A.~Albrecht and W.~H.~Zurek,
Nature Commun. \textbf{10}, no.1, 1024 (2019)
doi:10.1038/s41467-019-08426-4
[arXiv:1708.09353 [quant-ph]].


\bibitem{limousin}M.~Limousin, J.-P. Kneib, P. Natarajan, MNRAS, 356 (2005) 309.

\bibitem{Eliasdottir:2007md}
A.~Eliasdottir, M.~Limousin, J.~Richard, J.~Hjorth, J.~P.~Kneib, P.~Natarajan, K.~Pedersen, E.~Jullo and D.~Paraficz,
``Where is the matter in the Merging Cluster Abell 2218?,''
[arXiv:0710.5636 [astro-ph]].

\bibitem{r8}
A.~B.~Newman, T.~Treu, R.~S.~Ellis, D.~J.~Sand, C.~Nipoti, J.~Richard and E.~Jullo,
``The Density Profiles of Massive, Relaxed Galaxy Clusters: I. The Total Density Over 3 Decades in Radius,''
Astrophys. J. \textbf{765}, 24 (2013)
[arXiv:1209.1391 [astro-ph.CO]].
\bibitem{nfw} J. F. Navarro, C. S. Frenk, S. D. M. White, ``A universal density profile from hierarchical clustering,''  Astrophys. J. 
490:493 (1997).
\bibitem{r9}C. W. Misner,  K. S. Thorne,  J. A. Wheeler, Gravitation (Physics Series). Princeton University Press, 1973.
\bibitem{r7}R. D. Inverno, Introducing Einstein's Relativity (Clarendon Press, Oxford, 1992).
\bibitem{Aghanim:2018eyx}
N.~Aghanim \textit{et al.} [Planck],
``Planck 2018 results. VI. Cosmological parameters,''
Astron. Astrophys. \textbf{641}, A6 (2020)
[erratum: Astron. Astrophys. \textbf{652}, C4 (2021)]
[arXiv:1807.06209 [astro-ph.CO]].

\bibitem{r6}
A.~Matsumura,
``Field-induced entanglement in spatially superposed objects,''
Phys. Rev. D \textbf{104}, no.4, 046001 (2021)
[arXiv:2102.10792 [quant-ph]].






\end{thebibliography}
\end{document}